\documentclass[aps,prd,preprint]{revtex4}
\usepackage{graphicx,longtable}
\newcommand{\lw}[1]{\smash{\lower2.ex\hbox{#1}}}
\def\simlt{\rlap{\lower 3.5 pt\hbox{$\mathchar \sim$}}\raise 1pt \hbox {$<$}}
\def\simgt{\rlap{\lower 3.5 pt\hbox{$\mathchar \sim$}}\raise 1pt \hbox {$>$}}
\def\dfrac#1#2{\displaystyle\frac{#1}{#2}}

\newcommand{\nn}{\nonumber}
\newcommand{\beqa}{\begin{eqnarray}}
\newcommand{\eeqa}{\end{eqnarray}}
\newcommand{\tr}{{\rm tr}}
\newcommand{\ovl}[1]{\overline{#1}}
\newcommand{\wt}[1]{\widetilde{#1}}

\begin{document}
\draft

\title{Non-perturbative calculation of $Z_V$ and $Z_A$
in domain-wall QCD on a finite box}

\author{CP-PACS Collaboration:
  S.~Aoki\rlap,$^{\rm a}$
  M.~Fukugita\rlap,$^{\rm b}$
  N.~Ishizuka\rlap,$^{\rm a,c}$
  Y.~Iwasaki\rlap,$^{\rm a}$
  K.~Kanaya\rlap,$^{\rm a}$
  T.~Kaneko\rlap,$^{\rm d}$
  Y.~Kuramashi\rlap,$^{\rm d}$
  M.~Okawa\rlap,$^{\rm e}$
  Y.~Taniguchi\rlap,$^{\rm a}$
  A.~Ukawa$^{\rm a,c}$ and
  T.~Yoshi\'e$^{\rm a,c}$
}
\address{
$^{\rm a}$Institute of Physics,~University of Tsukuba,~Tsukuba,~Ibaraki 305-8571, Japan \\
$^{\rm b}$Institute for Cosmic Ray Research,
    University of Tokyo, Kashiwa, Chiba 277-8582, Japan \\
$^{\rm c}$Center for Computational Physics,
    University of Tsukuba, Tsukuba, Ibaraki 305-8577, Japan \\
$^{\rm d}$High Energy Accelerator Research Organization
    (KEK), Tsukuba, Ibaraki 305-0801, Japan \\
$^{\rm e}$ Department of Physics, Hiroshima University, Higashi-Hiroshima,
Hiroshima 739-8526, Japan\\
}

\date{\today}

\begin{abstract}
We report on a non-perturbative evaluation of the renormalization factors
for the vector and axial-vector currents, $Z_V$ and $Z_A$,
in the quenched domain-wall QCD (DWQCD) with plaquette 
and renormalization group improved gauge actions.
We take the Dirichlet boundary condition for both gauge and
domain-wall fermion fields on the finite box,
and  introduce the flavor-chiral Ward-Takahashi identities
to calculate the renormalization factors.
As a test of the method, we numerically confirm the expected relation that 
$Z_V \simeq Z_A$ in DWQCD.
Employing two different box sizes for the numerical simulations
at several values of the gauge coupling constant $g^2$ and 
the domain-wall height $M$,
we extrapolate $Z_V$ to the infinite volume to remove $a/L$ errors.
We finally give the interpolation formula of $Z_V$ in the infinite volume
as a function of $g^2$ and $M$.

\end{abstract}


\maketitle

\section{Introduction}
\label{sec:intro}
Recent lattice calculations in the domain-wall QCD (DWQCD) have shown that 
the good chiral property of domain-wall fermions
leads to a good scaling behavior of physical observables
such as quark masses and $B_K$\cite{cppacs_bk}.
Aside from the quenched approximation,
the use of perturbative renormalization factors 
is the largest source of the uncontrolled systematic 
errors in these calculations.   Some kind of the non-perturbative renormalization 
is required to reduce the total error to a few percent level except 
that from the quenched approximation.

There exist two popular methods for the non-perturbative renormalization 
in lattice QCD: one is the RI-MOM (Regularization Independent MOMentum
subtraction) scheme\cite{RIMOM}, 
and the other is the SF (Schr\"odinger Functional) scheme\cite{alphaSF}.
The former method is simpler and has already been applied to DWQCD\cite{RBC_Z}.
The latter one is more suitable to evaluate the scale dependent renormalization
factors. It is rather complicated, however, to implement the SF scheme 
in DWQCD.

In this paper we formulate the finite volume method very similar to the
SF scheme, to calculate the scale independent renormalization factors,
$Z_V$ and $Z_A$. We employ the SF boundary condition for the gauge fields,
equivalent to the Dirichlet boundary condition in the absence of the boundary 
fields, while the boundary quark fields with the simple Dirichlet boundary 
condition, which is different from the  SF boundary condition for quarks,
are introduced to construct the gauge invariant observables.
In the case of the scale dependent renormalization factors such as $Z_P$,
an extra perturbative calculation is required to convert the renormalization
factors calculated in some scheme with the special boundary condition( the 
ours or the SF) to the one defined in the conventional $\overline{\rm MS}$ 
scheme. 
In the case of $Z_V$ or $Z_A$, however,
the same renormalization factors are obtained from different boundary
conditions,
since the flavor-chiral Ward-Takahashi identities uniquely
determine them.

Using the finite volume method, we can calculate $Z_V$ and $Z_A$ 
non-perturbatively at the massless point, so that the systematic error
associated with the chiral extrapolation can be removed.
The calculation of the scale independent renormalization factors 
for vector and axial-vector currents is the first step to the calculation of 
the scale-dependent renormalization factors for the quark mass and $B_K$.
In addition to this purpose 
we can use our calculations to probe the chiral symmetry
in DWQCD. For example, since the chiral symmetry predicts $Z_V = Z_A$,
the difference of the two renormalization factors can be used to measure
the size of the chiral symmetry breaking in DWQCD.

This paper is organized as follows.
In sect.~\ref{sec:finite} we formulate DWQCD on a finite box.
In particular we give a detailed description of the quark boundary conditions
and explicit forms for the correlation functions which includes the boundary
quark fields.
In sect.~\ref{sec:WT} utilizing the vector an axial-vector Ward-Takahashi
identities, we introduce the conditions which determine the renormalization factors
for the vector and axial-vector currents. We explicitly give the renormalization 
factors $Z_V$ and $Z_A$ in terms of the correlation functions on the finite box.
In sect.~\ref{sec:test} we present results of numerical tests for our method.
By investigating the behavior of the quark mass defined through the axial
Ward-Takahashi identity as a function of the time, we show that the effect of
the Dirichlet boundaries to the zero modes rapidly disappear
away from the boundaries. We also show that the expected relation
$Z_V\simeq Z_A$ is satisfied for the sufficiently large $N_s$, the size of the
5th dimension of DWQCD.
In sect.~\ref{sec:result} we calculate the renormalization factors
at several values of the gauge coupling constant $g^2$ and the domain-wall height
$M$ for both plaquette and renormalization group(RG) improved gauge actions
in the quenched approximation. Using data from two different lattice volumes
we extrapolate $Z_V$ to the infinite volume, in order to remove 
possible $O(a/L)$ errors. We globally fit $Z_V$ in the infinite volume
as a function of $g^2$ and $M$.
Our conclusion and discussion are given in sec.~\ref{sec:concl}.

\section{DWQCD on a Finite Box}
\label{sec:finite}

\subsection{Gauge action}

The gauge action is give by
\[
S[U] = \frac{2}{g_0^2}\{ c_0 \sum_P {\rm Re} \tr (I-U_P) + c_1 \sum_R {\rm Re}
\tr (I-U_R) \}
\]
where $U_P$ is the product of the gauge link variables along the plaquette loop $P$
and $U_R$ is the one along the rectangular loop $R$, with the
normalization $c_0+8 c_1 = 1$. Note that the action with $c_1=0$ corresponds
to the plaquette action and $c_1=-0.331$ is the RG improved one obtained by
Iwasaki\cite{iwasaki}.

In the finite volume scheme such as the SF scheme,
the theory is defined on  $L^3\times T$ lattice with cylinder geometry,
i.e. the periodic-type boundary condition (PBC) in the spatial directions and 
the Dirichlet boundary condition (DBC) in the time direction.
Throughout this paper the convention that $L = N_l a$ and
$T = N_t a$ is used. In this case
the dynamical variable are
$U(x)_k$ with times $x_0 = a,\dots , T-a$ and $U(x)_0$ with
$x_0 = 0,\dots , T-a$ (i.e. inside the cylinder), while
Dirichlet boundary conditions are imposed on the fields $U(x)_k$ at
$x_0 = 0$ and $T$ as follows:
\[
U(\vec{x},x_0=0)_k = \exp [a C_k], \qquad 
U(\vec{x},x_0=T)_k = \exp [a C_k^\prime],
\]
where $C_k$ and $C_k^\prime$ are diagonal matrices\cite{alphaSF}:
\[
(C_k)_{ij} = \phi_i \delta_{ij}, \qquad (C_k^\prime)_{ij}= \phi_i^\prime
\delta_{ij} .
\]
In the calculation of renormalization factors, we take
$\phi_i =\phi^\prime_i = 0$; zero boundary fields.

In order to remove $O(a)$ errors caused by the DBC, we modify the
weight $c_0$ and $c_1$ in the action near the boundaries.
In the case of the plaquette action the perturbative calculation gives
\[
c_0\rightarrow c_t = 1-0.089 g_0^2 - 0.030g_0^4 +O(g_0^6)
\]
for each time-space plaquette $P_{0k}$ which just touches one of the boundaries.
(The time coordinate for the center of the plaquette $x_0 = a/2$ or $T-a/2$.)
In the case of the RG action, there exist several choices but we adopt
the following one which remove $O(a)$ term at the tree level\cite{TAI}: 
\[
c_1 \rightarrow \frac{3}{2} c_1 
\]
for each time-space-space rectangle $R_{0kk}$ which has 
exactly 2 lines on a boundary. (Again 
the time coordinate for the center of the rectangle $x_0 = a/2$ or $T-a/2$.)
A proof for the $O(a)$ improvement by this choice is given in Ref.\cite{AFW}.

\subsection{Domain-wall fermion on a finite box}

The domain-wall fermion action is given by\cite{shamir}
\[
S_F = \bar\psi (x,s) D(x,s;y,t) \psi (y,t)
\]
where $x,y$ are 4 dimensional coordinates and $s,t$ are coordinates in the
fifth dimension, which run from 1 to $N_s$. 
For the short-handed notation, $X=(x,s)$ and $Y=(y,t)$ is used. Explicitly
\begin{eqnarray*}
D(X,Y)&=& (5-M)\delta_{XY}-D^4(x,y)\delta_{st}-D^5(s,t)\delta_{xy} \\
D^4(x,y)&=&P_{-\mu} U(x)_\mu \delta_{y,x+\hat\mu a}
        +P_\mu U(y)_\mu^\dagger\delta_{y,x-\hat\mu a} , \\
D^5(s,t) &=& \left\{
\begin{array}{ll}
P_L \delta_{t,s+1}- m_f P_R \delta_{s,N_s} & (s=1) \\
P_L\delta_{t,s+1}+P_R\delta_{t,s-1} & (1 < s <N_s) \\
-m_f P_L\delta_{t,1}+P_R\delta_{t,s-1} & (s=N_s) \\
\end{array} \right. \\
P_{\pm \mu} &=&\dfrac{1}{2}(1\pm\gamma_\mu), \qquad
P_{R/L} =\dfrac{1}{2}(1\pm\gamma_5) .
\end{eqnarray*}
The quark field is defined as usual:
\beqa
q(x)&=& P_L \psi(x,1)+ P_R\psi(x,N_s) \equiv P_L(s)\psi(x,s) \nn \\
\bar q(x) &=& \bar\psi (x,N_s)P_L + \bar\psi(x,1)P_R \equiv \nn
\bar\psi (x,s) P_R(s)
\eeqa
where $P_L(s) = P_L\delta_{s,1}+P_R\delta_{s,N_s}$ and
$P_R(s) =P_R\delta_{s,1}+P_L\delta_{s,N_s}$.
The following property is useful:
\[
[D(x,s;y,t)]^\dagger = \gamma_5 D(y,s^p;x,t^p) \gamma_5
\]
where $s^p = N_s+1-s$, and the dagger here is applied only to color
and spinor indices. It is explicitly given by
\[
\ovl{D(x,s;y,t)_{\beta\alpha}^{ba}} = 
[\gamma_5 D(y,s^p;x,t^p)^{ab} \gamma_5]_{\alpha\beta}
\]
with  color indices $a,b$ and spinor indices $\alpha,\beta$.

In the finite volume scheme we may rewrite it as
\beqa
S_F &=& \sum_{a < x_0, y_0 < T} \bar\psi (x,s) D(x,s;y,t) \psi (y,t) 
+ O(\bar\rho\rho, \bar\rho^\prime\rho^\prime)\nn \\
&-&\wt{c_t}\sum_{x_0=0}\left[\bar\psi(x,s)P_- U(x)_0 \psi(x+\hat 0 a,s) 
+\bar\psi(x+\hat 0 a,s)P_+U^\dagger(x)_0\psi(x,s)\right] \nn \\
&-&\wt{c_t}\sum_{x_0=T}\left[\bar\psi(x-\hat 0 a,s)
P_- U(x-\hat 0 a)_0 \psi(x,s) 
+\bar\psi(x,s)P_+U^\dagger(x-\hat 0 a)_0\psi(x-\hat 0 a,s)\right] 
\label{eq:dwSF}
\eeqa
with the coefficient $\wt{c_t}$ for the boundary counter term
and the boundary condition that
\beqa
\psi(\vec{x},x_0=0,s) = P_+ P_L(s)\rho(\vec{x}), & \qquad &
\bar\psi(\vec{x},x_0=0,s) =\bar\rho(\vec{x}) P_R(s)P_- \nn \\ 
\psi(\vec{x},x_0=T,s) = P_- P_L(s)\rho^\prime(\vec{x}), &\qquad&
\bar\psi(\vec{x},x_0=T,s) =\bar\rho^\prime(\vec{x}) P_R(s)P_+ \nn
\eeqa
where $P_\pm = P_{\pm 0}$. 
Terms which contain two external fields are not explicitly written 
in the first line in eq.~(\ref{eq:dwSF}), since they do not 
contribute to the correlation functions we are interested in.
Note that this boundary condition is different from the
SF boundary condition for quarks\cite{sint}, since this condition is invariant
under the chiral transformation of the domain-wall fermion defined by
\beqa
\psi(x,s) &\rightarrow & e^{i w(s)}\psi(x,s) \\
\bar\psi(x,s) &\rightarrow & \bar\psi(x,s) e^{-i w(s)} 
\eeqa
with $w(s)=\theta (s-(N_s+1)/2)$ while
the SF boundary condition for quarks must break the chiral symmetry\cite{sint}.
In the continuum limit the boundary terms in the latter case becomes
\beqa
\bar \psi(\vec{x},0) P_-\psi(\vec{x},0)+
\bar \psi(\vec{x},T) P_+\psi(\vec{x},T) ,
\eeqa
which manifestly breaks the chiral symmetry.
It may be possible to formulate the domain-wall fermion which satisfies 
the corresponding SF boundary condition on a cylinder\cite{AKT}. 

The classical solution which satisfies the Dirac equation
\beqa
D(X,Y)\psi_{cl}(Y) &=& 0 ,\qquad 0 < x_0 < T
\label{eq:dirac}
\eeqa
with boundary values
\beqa
\psi_{cl}(X)\vert_{x_0=0}&=& P_+ P_L(s)\rho(\vec{x}), \quad
\psi_{cl}(X)\vert_{x_0=T}= P_- P_L(s)\rho^\prime (\vec{x}),
\label{eq:bvalue}
\eeqa
is given by\cite{LW} 
\beqa
\psi_{cl} (X) &=&\wt{c_t} \sum_Y S(X,Y)\left[
U(y-\hat 0 a)_0^\dagger P_+ P_L(t)\rho(\vec{y})\vert_{y_0=a}
+U(y)_0P_-P_L(t)\rho^\prime(\vec{y})\vert_{y_0=T-a}
\right]
\eeqa
where $S(X,Y)$ is the propagator with the zero boundary value:
\beqa
D(X,Y) S(Y,Z) &=& \delta_{X,Z}, \qquad 0 < x_0 < T \\
P_+ S(X,Y)\vert_{x_0=0} &=& P_-S(X,Y)\vert_{x_0=T} = 
 S(X,Y)P_-\vert_{y_0=0} = S(X,Y)P_+\vert_{y_0=T} = 0 . \nn
\eeqa
Note that the above expression for $\psi_{cl}$ is not valid at $x_0=0$ or $T$.
To show eq.~(\ref{eq:dirac}), it is enough to see
\beqa
D(X,Y)\psi_{cl}(Y) &=& \wt{c_t}\sum_{Y, 0<y_0< T}D(X,Y)\psi_{cl}(Y)
+\wt{c_t}\sum_{Y, y_0=0,T}D(X,Y)\psi_{cl}(Y) \nn \\
&=& \wt{c_t}\left[\delta_{x_0,a}
U(x-\hat 0 a)_0^\dagger P_+ P_L(s)\rho(\vec{x})+\delta_{x_0,T-a}
U(x)_0 P_- P_L(s)\rho^\prime(\vec{x})\vert_{x_0=T-a} \right]\nn \\
&-& \wt{c_t}\left[ \delta_{x_0,a}P_+U(x-\hat 0 a)_0^\dagger\psi_{cl}(\vec{x},0,s)
+\delta_{x_0,T-a} P_-U(x)_0\psi_{cl}(\vec{x},T,s) \right] = 0 \nn
\eeqa
for $0 < x_0 < T$ with boundary values (\ref{eq:bvalue}).
In the actual simulations,
the propagator $S(X,Y)$ can be easily obtained by 
solving the Dirac equation numerically with the condition that
$U(\vec{x},x_0=0)_0 = U(\vec{x},x_0=T-a)_0 = 0$.

Now let us consider the path-integral for the fermion with 
source $\eta(x)$, $\bar\eta(x)$ and the boundary fields
$\rho$, $\bar\rho$, $\rho^\prime$, $\bar\rho^\prime$:
\beqa
Z_F(\eta,\bar\eta,\rho,\bar\rho,\rho^\prime,\bar\rho^\prime)
&=& \int {\cal D}\psi {\cal D}\bar\psi \exp [-S_F +
\bar\psi(x,s)P_R(s)\eta(x)+\bar\eta(x)P_L(s)\psi(x,s)]
\eeqa
To perform path-integral, we introduce the following change of variables
\[
\psi(x,s)=\psi_{cl}(x,s)+\chi(x,s)\qquad
\bar\psi(x,s)=\bar\psi_{cl}(x,s)+\bar\chi(x,s)\qquad ,
\]
with the boundary condition that $\chi (x,s) = \bar\chi(x,s) = 0 $
at $x_0 =0$ and $T$. 
Integrating out $\chi$ and $\bar\chi$ and 
using the fact that the classical background fields $\psi_{cl}$ and 
$\bar\psi_{cl}$ satisfy Dirac equation except boundaries,
one finally obtain
\beqa
Z_F &=& \det D \exp [ -\bar\psi_{cl}(X) D(X,Y)\psi_{cl}(Y)
+\bar\eta (x) P_L(s) S(X,Y) P_R(t)\eta(y)  \nn \\
\nn \\
&+&\bar\psi_{cl}(x,s)P_R(s)\eta(x)+\bar\eta(x)P_L(s)\psi_{cl}(x,s) ] .
\eeqa

Introducing the boundary fields as
\[
\zeta (x) = \dfrac{\overrightarrow \delta}{\delta \bar\rho(x)},
\qquad
\bar\zeta (x) = \dfrac{\overleftarrow \delta}{\delta \rho(x)},
\]
and denoting $q(x)=P_L(s)\psi(x,s)$ and $\bar q(x)=\bar\psi(x,s)P_R(s)$,
we list all fermionic correlation functions used in this report as follows.
\beqa
\langle q(x) \bar q(y) \rangle & = &
\dfrac{\overrightarrow \delta}{\delta \bar\eta(x)}
\log Z_F
\dfrac{\overleftarrow \delta}{\delta \eta(y)}
= P_L(s) S(X,Y)P_R(t) \nn \\
\langle q(x) \bar\zeta (y)\rangle &=&
\dfrac{\overrightarrow \delta}{\delta \bar\eta(x)}
\log Z_F
\dfrac{\overleftarrow \delta}{\delta \rho (y)}
= P_L(s) S(X,Y)U(y-\hat 0 a)_0^\dagger P_+ P_L(t)\vert_{y_0=a} \nn \\
\langle \zeta(x) \bar q(y)\rangle &=&
P_R(s) P_- U(y-\hat 0 a)_0 S(X,Y)  P_R(t)\vert_{x_0=a} 
= \gamma_5 \langle q(y) \bar\zeta (x)\rangle^\dagger \gamma_5 \nn \\
\langle q(x) \bar\zeta^\prime (y)\rangle &=&
P_L(s) S(X,Y)U(y)_0 P_- P_L(t)\vert_{y_0=T-a} \nn \\
\langle \zeta^\prime (x) \bar q(y)\rangle &=&
= P_R(s) P_+ U(y)_0^\dagger S(X,Y)  P_R(t)\vert_{x_0=T-a} 
\gamma_5 \langle q(y) \bar\zeta^\prime (x)\rangle^\dagger \gamma_5\\
\langle \zeta^\prime (x) \bar\zeta (y)\rangle &=&
\dfrac{\overrightarrow \delta}{\delta \bar\rho^\prime(x)}
\log Z_F
\dfrac{\overleftarrow \delta}{\delta \rho (y)}
= P_R(s) P_+ U(x)_0^\dagger S(X,Y)U(y-\hat 0 a)_0^\dagger P_+ P_L(t)
\vert_{x_0=T-a,y_0=a} \nn \\
\langle \zeta (x) \bar\zeta^\prime (y)\rangle &=&
P_R(s) P_- U(x-\hat 0 a)_0 S(X,Y)U(y)_0 P_- P_L(t)
\vert_{x_0=a,y_0=T-a} 
= \gamma_5 \langle \zeta^\prime (y) \bar\zeta (x)\rangle^\dagger \gamma_5 \nn
\eeqa
where $^\dagger$ is applied to only color and flavor indices. 

%
It is finally noted that the twisted boundary condition in
the spatial directions can be imposed for the quarks\cite{alphaSF}, by replacing
\[
U(x)_k \longrightarrow \lambda_k U(x)_k
\]
where $ \lambda_k = e^{i a \theta/L}$

\section{Determination of renormalization factors}
\label{sec:WT}

\subsection{Ward-Takahashi identities}

The integrated version of Ward-Takahashi(WT) identities are used
to determine renormalization factors for vector and axial-vector currents,
$Z_V$ and $Z_A$\cite{alphaSF}. 
Let $R$ be a space-time region with smooth boundary $\partial R$, and
${\cal O}_{\rm int}$ and ${\cal O}_{\rm ext}$ are observables localized in the
interior and the exterior of $R$ respectively. The vector WT identity
reads
\[
\int_{\partial R} d\sigma_\mu (x) \langle V_\mu^a (x) {\cal O}_{\rm int}
{\cal O}_{\rm ext} \rangle  =  - \langle (\delta_V^a {\cal O}_{\rm int})
{\cal O}_{\rm ext} \rangle 
\]
while the axial-vector WT becomes
\[
\int_{\partial R} d\sigma_\mu (x) \langle A_\mu^a (x) {\cal O}_{\rm int}
{\cal O}_{\rm ext} \rangle =  - \langle (\delta_A^a {\cal O}_{\rm int})
{\cal O}_{\rm ext} \rangle + 2m \int_R d^4x \langle P^a(x){\cal O}_{\rm int}
{\cal O}_{\rm ext}\rangle 
\]
where $V_\mu^a$ ($A_\mu^a$) is the (axial-)vector current and
$P^a$ is the pseudo-scalar density:
\beqa
V_\mu^a (x) &=& \bar q(x) \gamma_\mu \dfrac{\tau^a}{2} q(x),\quad
A_\mu^a (x) = \bar q(x) \gamma_\mu\gamma_5 \dfrac{\tau^a}{2} q(x),\quad
P^a (x) = \bar q(x) \gamma_5 \dfrac{\tau^a}{2} q(x) . \nn
\eeqa

\subsection{Vector current}
We take $R = L^3 \times (0, x_0) $, so that $\partial R $ is consist of 
3-dimensional spaces at $t=0$ and at $t=x_0$. 
As a gauge invariant observable, we choose
${\cal O}_{\rm ext}={\cal O}^{\prime a}$ and 
${\cal O}_{\rm int} ={\cal O}^a$ with
\beqa
{\cal O}^a &=& a^6 \sum_{\vec{u},\vec{v}}\bar\zeta (u) \gamma_5
\dfrac{1}{2}\tau^a \zeta(v), \quad
{\cal O}^{\prime a} = a^6 \sum_{\vec{u},\vec{v}}\bar\zeta^\prime (u) \gamma_5
\dfrac{1}{2}\tau^a \zeta^\prime (v),  \nn
\eeqa
where $\tau^a$ is the Pauli matrix for flavors with 
$\tr \tau^a\tau^b = 2 \delta_{ab}$ and $\tau^a\tau^b = \delta^{ab} 
+ i \epsilon^{abc}\tau^b\tau^c$, and $\zeta$ and 
$\zeta^\prime$ correspond to our boundary fields.
With this choice and 
\[
\delta_V^a {\cal O}^b = - i\epsilon^{abc}{\cal O}^c ,
\]
the vector WT identity gives the relation that
\beqa
Z_V (1+ b_V m_q a) f_V(x_0) &=& f_1 
\label{eq:zv}
\eeqa
where
\beqa
f_V(x_0) &=& \dfrac{a^3}{6L^6}\sum_{\vec{x}} i\epsilon^{abc}
\langle {\cal O}^{\prime a} V_0^b (x) {\cal O}^c\rangle \\
f_1 &=& -\dfrac{1}{3 L^3} \langle {\cal O}^{\prime a} {\cal O}^a \rangle .
\eeqa
From eq.~(\ref{eq:zv}) we can determine
$Z_V$, the renormalization factor for the vector current,
together with the $b_V$, one of the $O(ma)$ improvement coefficients.
Note that $b_V=0$ if the chiral symmetry of DWQCD is exactly satisfied.

\subsection{Axial-vector current}
For the axial-vector current, we take $R = L^3\times (y_0-t, y_0+t)$,
${\cal O}_{\rm int} = A_0^b (y_0)$ and
\[
{\cal O}_{\rm ext} = -\epsilon^{cde} {\cal O}^{\prime d}{\cal O}^e,
\]
and plug them into the axial-vector WT identity with $m = 0$.
We then obtain 
\beqa
Z_A^2 f_{AA}^I (y_0, x_0^+,x_0^-) &=& 2 Z_V f_V(y_0) = 2 f_1
\label{eq:za1}
\eeqa
where
\beqa
f_{AA}^I (y_0, x_0^+,x_0^-) &=& -\dfrac{a^6}{6L^6}
\sum_{\vec{x},\vec{y}}\epsilon^{abc}\epsilon^{cde}
\langle {\cal O}^{\prime d}\{A_0^a(x_0^+,\vec{x})-A_0^a(x_0^-,\vec{x})\}
A_0^b(y){\cal O}^e\rangle
\eeqa
with $x_0^\pm = y_0 \pm t$. 

We finally define the quark mass $m_{\rm AWTI}$ through the following
WT identity:
\beqa
f_A(x_0+a)-f_A(x_0-a) &=& 4 (m_{\rm AWTI} a) f_P(x_0)
\eeqa
where
\beqa
f_A(x_0) &=& - \dfrac{a^6}{3}\langle A_0^a(x){\cal O}^a\rangle, \qquad
f_P(x_0) = - \dfrac{a^6}{3}\langle P^a(x){\cal O}^a\rangle .
\eeqa

\section{Test of the formulation by numerical simulations}
\label{sec:test}

\subsection{Effects of boundaries to quark masses}
\begin{figure}[p]
\centering{\includegraphics[width=115mm,angle=0]{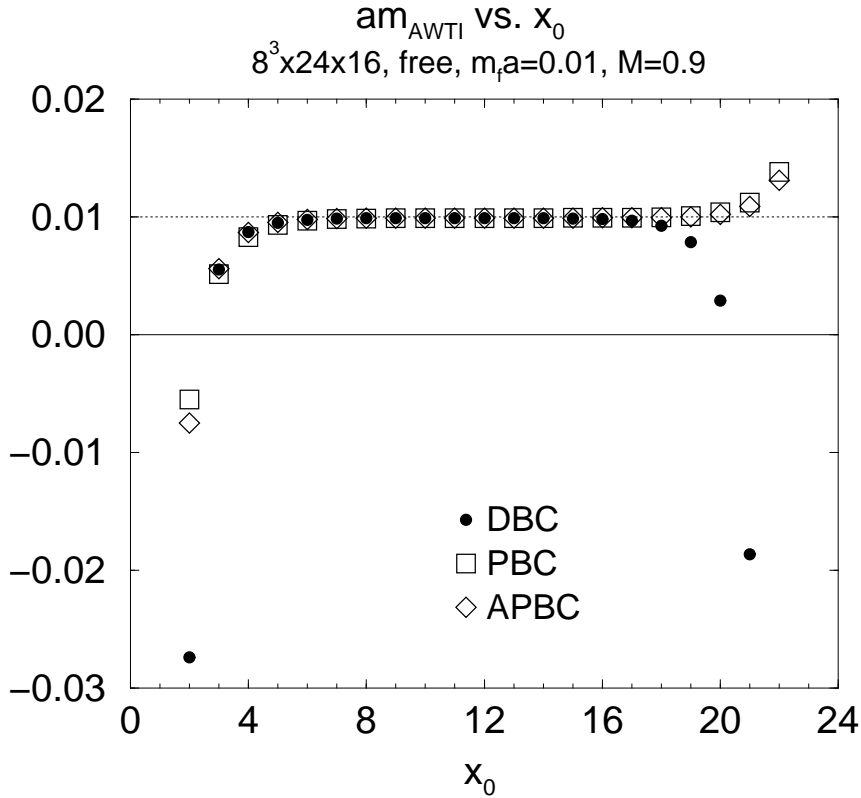}}
\vspace{-0.5cm}
\caption{$a m_{\rm AWTI}$ as a function of $x_0$ with 
Dirichlet(solid circles),
periodic(open squares) and anti-periodic(open diamonds) 
boundary conditions.}
\label{fig:mAWTI}
\end{figure}

Since the boundary condition in time with $\rho=\bar\rho = 0$ is 
identical to the Shamir's domain-wall(Dirichlet) boundary 
condition\cite{shamir},  
extra zero modes may appear near $x_0 = 0$ and $T$.
One has to check whether these unwanted zero modes induce an extra 
contribution to the low energy observables at $ 0 \ll x_0 \ll T$.
Here we consider the quark mass, $a m_{\rm AWTI}$, defined through 
the axial Ward-Takahashi identity(AWTI).
In Fig.~\ref{fig:mAWTI},
we plot $ a m_{\rm AWTI}$ for free theory as a function of $x_0$ with
Dirichlet, periodic and anti-periodic boundary conditions at the bare
quark mass $m_f a$=0.01, on an $8^3\times 24 \times 16$ lattice,
with the domain-wall height $M$=0.9.
The dependence of $a m_{\rm AWTI}$ on the boundary condition,
which is visible near the boundaries, disappears away from them.
Therefore we conclude at least for the free case
that the extra zero modes associated with 
the Dirichlet boundary condition gives negligible effects 
to the determination of the renormalization 
factors evaluated at $x_0 \simeq T/2$.

\begin{figure}[p]
\centering{\includegraphics[width=105mm,angle=0]{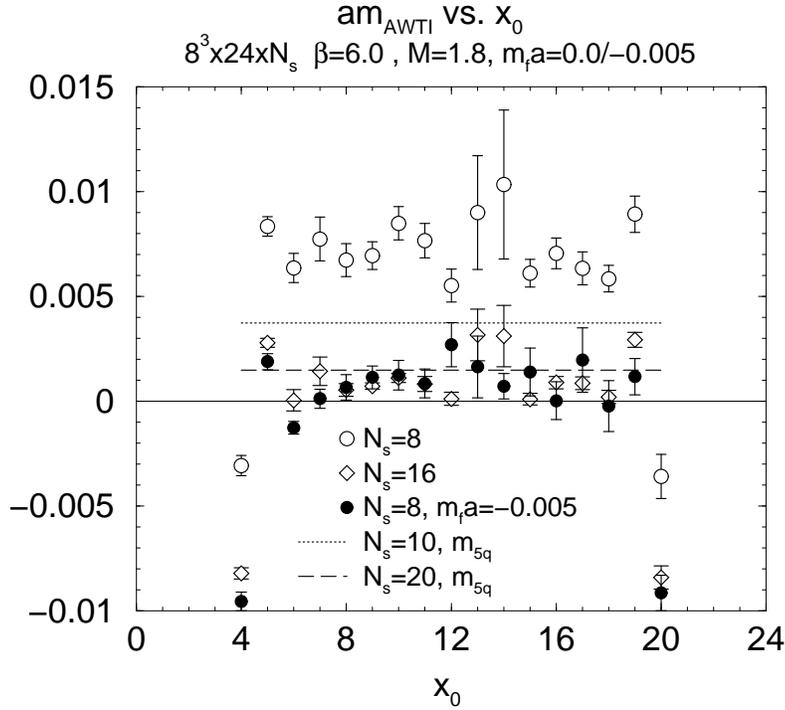}}
\vspace{-0.5cm}
\caption{$a m_{\rm AWTI}$ as a function of $x_0$ at $\beta\!=\!6.0$ on 
an $8^3\!\times\!24\!\times\!N_s$ lattice at 
$m_f a=0$ with $N_s=8$(open circles) and 16(open diamonds)
and at $m_f=-0.005$ with $N_s=8$(solid circles), 
together with $m_{5q}$ at $N_s=10$(dotted line) and $N_s=20$
(dashed line)\protect{\cite{chiral}}.}
\label{fig:mq}
\end{figure}

In Fig.~\ref{fig:mq}, $ a m_{\rm AWTI}$ in the quenched DWQCD
with our boundary condition is plotted as a function of $x_0$
on $8^3\times 24\times N_s$ lattices with $m_f a=0$ and $M=1.8$ 
at $\beta = 6.0$ for the plaquette gauge action. 
Since the $x_0$ dependence is weak away from the boundaries,
we non-perturbatively confirm the conclusion in the free case
that the effect of the Dirichlet boundary condition is negligible.
Interestingly $m_{\rm AWTI}$ is non-zero even at $m_f a = 0$, 
and becomes smaller for larger $N_s$. Moreover
the value is consistent with $m_{5q}$,
a measure of the explicit chiral symmetry breaking
calculated from the conserved axial-vector current of DWQCD\cite{chiral}.
This fact suggests
$m_{\rm AWTI}$ in our finite volume scheme may be a better alternative 
as the measure of explicit chiral symmetry breaking in DWQCD, 
since it can be calculated directly at $m_f a = 0$ with much less
computational cost.
Note also that
the large explicit breaking in $m_{\rm AWTI}$ at $N_s=8$ 
(open circles) is compensated if one takes a negative quark mass 
of $m_f=-0.005$ (filled circles).  This demonstrates that
the domain-wall fermion at $N_s\not=\infty$ 
can be considered as a highly improved Wilson fermion\cite{vpp}.

\subsection{Renormalization factors}
\begin{figure}[p]
\centering{\includegraphics[width=105mm,angle=0]{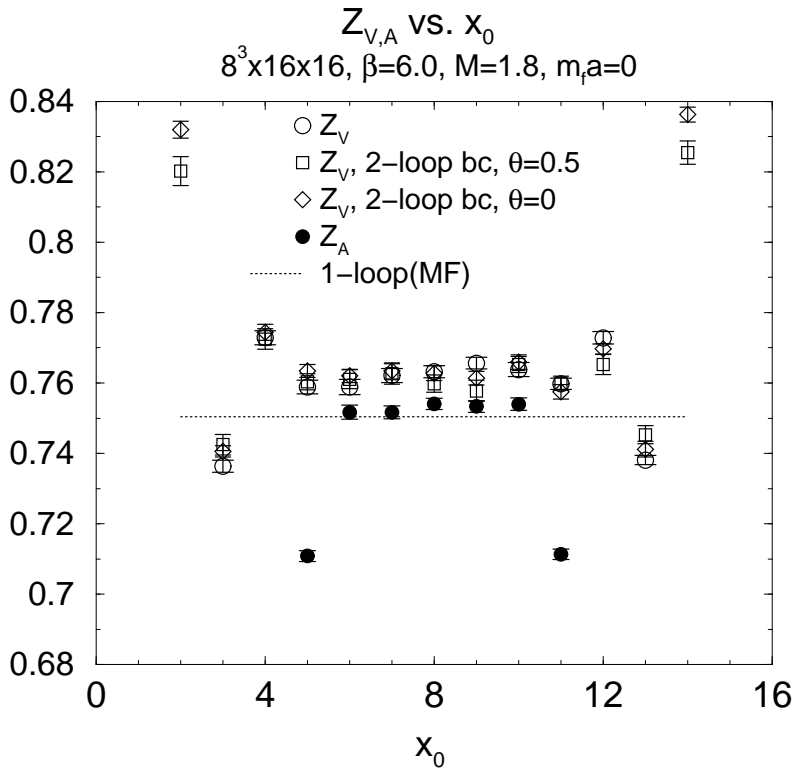}}
\vspace{-0.5cm}
\caption{$Z_{V}$ and $Z_A$ as a function of $x_0$ at $\beta=6.0$ on 
an  $8^3\times 16\times 16$ lattice with $M=1.8$ and $m_f a = 0$.
We compare the results from the boundary counter-terms 
at tree-level(circles) with those at 2-loop(squares and diamonds) as well
as those at $\theta = 0$ with that at $\theta = 0.5$(squares).}
\label{fig:zVA}
\end{figure}

The non-perturbative renormalization factors for vector and axial-vector 
currents are defined by
$Z_V (1 + b_V m_f a) = f_1/ f_V (x_0) $ and
$Z_A^2 = 2f_1/f_{AA}(x_0, x_0^+, x_0^-)$,
where we fix $x_0^\pm = T/2\pm T/4$ and
put $ m_f a = 0$ for $Z_A$.
In Fig.~\ref{fig:zVA}, $Z_{V}$ and $Z_A$ are plotted as a function of $x_0$
on $8^3\times 16\times 16$ at $\beta = 6.0$ with $M=1.8$ and $m_f=0$.
Similar to the case of $ a m_{\rm AWTI}$
a plateau is seen away from the boundaries. 
The relation $Z_V = Z_A$, valid exactly in perturbation theory,
is satisfied non-perturbatively within 1--2\%\footnote{
Note however that a small difference between $Z_V$ and $Z_A$
is beyond the statistical errors. This small difference is expected to vanish
exponentially as $N_s\rightarrow\infty$.}.
Moreover the magnitude of $Z_{V,A}$ almost 
agrees with the mean-field(MF) improved one-loop value.
We also observe that $Z_V$ is insensitive to boundary parameters such as
the 2-loop boundary counter-terms for gauge fields and
the parameter $\theta$ of the twisted boundary condition for quarks.

\subsection{Dependence of $Z_{V,A}$ on $M$}
\begin{figure}[p]
\centering{\includegraphics[width=120mm,angle=0]{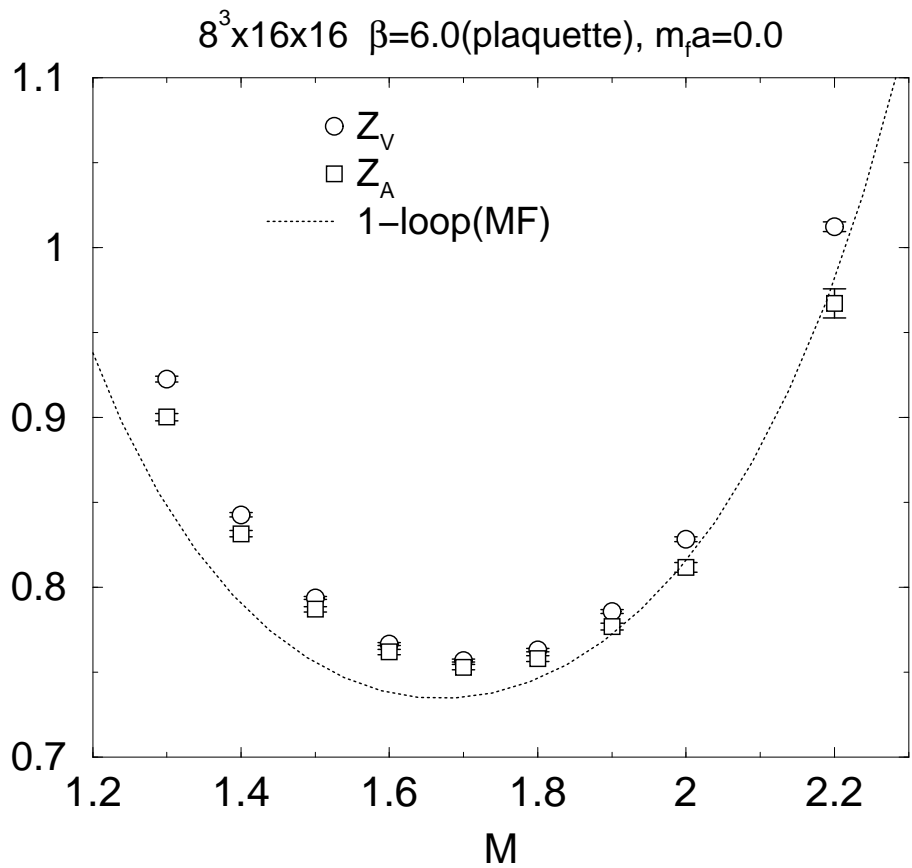}}
\centering{\includegraphics[width=120mm,angle=0]{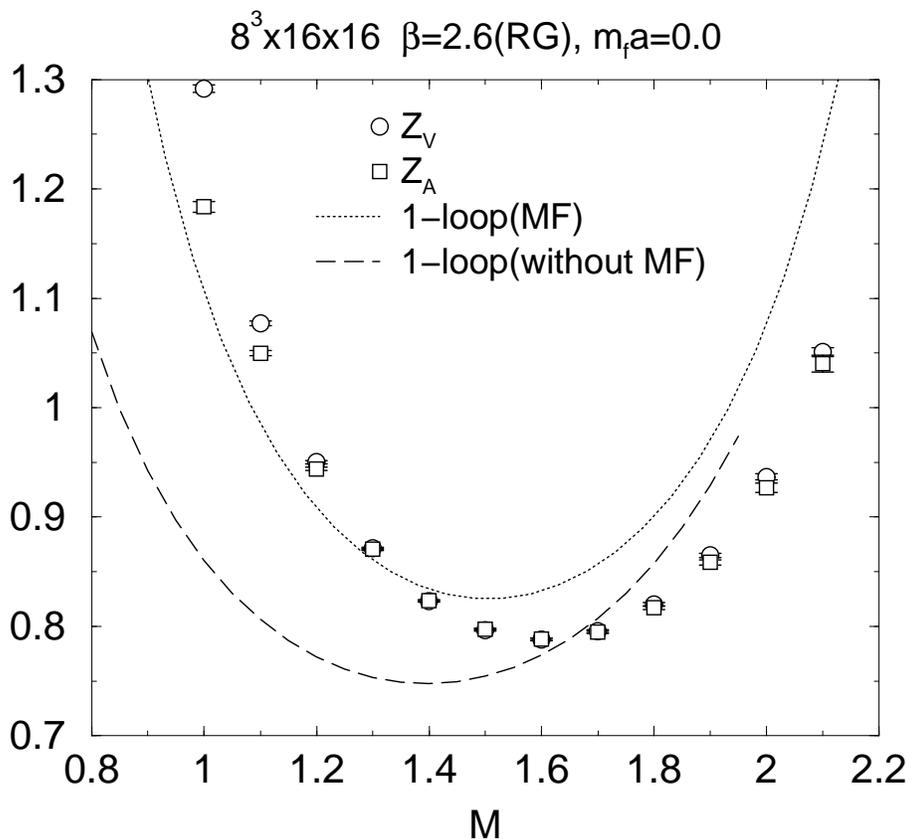}}
\caption{$Z_{V}$ and $Z_A$ vs $M$ on $8^3\times 16\times 16$ 
at $\beta=6.0$ for plaquette action(upper) and at $\beta=2.6$
for the RG action(lower). Perturbative estimates are given at 1-loop
with the MF improvement (solid lines) and without it (dashed line).
}
\label{fig:result}
\end{figure}

We calculate $Z_{V}$  and $Z_A$ in the quenched DWQCD at 
$a^{-1} \simeq 2$ GeV with the plaquette action ($\beta = 6.0$) and 
with the renormalization group(RG) improved action ($\beta = 2.6$)
on an $8^3\times 16 \times 16$ lattice with $m_f a = 0$ for $M=1.0\sim 2.2$.
The results are summarized in Fig.~\ref{fig:result},
where $Z_V$ and $Z_A$ are plotted as a function of $M$, 
together with one-loop perturbative estimates with and without 
mean-field (MF) improvement\cite{AIKT}.
For both gauge actions, $Z_V \simeq Z_A$ holds, and they have a minimum 
at $M\simeq 1.7$ for the plaquette action
or $M\simeq 1.6$ for the RG action.
The deviation from $Z_V = Z_A$ becomes larger as $M$ goes far away
from the minimum. This suggests that the chiral symmetry breaking 
effect is proportional to $ \vert M-M_{\rm min.}\vert^{N_s}$.
Perturbative estimates without MF improvement 
fail, particularly for the plaquette action for which 
the curve can not be placed in the figure.
The MF improvement makes the agreement much better for both actions.

\section{Results}
\label{sec:result}

We extract $Z_V$ and $Z_A$ at various values of 
$g^2$ for both plaquette and RG improved gauge actions 
on a  $N_l^3\times N_t\times N_s = 8^3\times 16\times 16$ 
lattice. In addition we employ a different 4 dimensional lattice size, 
$N_l^3\times N_t = 12^3\times 24$ or $4^3\times 8$, while keeping $N_s=16$, 
in order to investigate $a/L =1/N_l$ dependences of $Z_V$ and $Z_A$.
Simulation parameters are given in Table~\ref{tab:param}, 
together with the lattice spacing $a$, obtained from
the global parametrization for the string tension as a function of $g^2$:
\beqa
\sigma^{1/2} a &=& a(g^2)\frac{1 + c_2 \hat a^2(g^2)+c_4 \hat a^4(g^2)
+c_6 \hat a^6(g^2)}{c_0} \\
a(g^2) &=& (b_0 g^2)^{-b_1} \exp[-\frac{1}{2b_0 g^2}] \\
\hat a(g^2) &=&\frac{a(g^2)}{a(g_0^2)}
\eeqa
where  $b_0 = 11/(4\pi)^2$, $b_1 = 102/(4\pi)^4$ (the coefficients of $\beta$ 
function 
in the quenched theory). The coefficients of the parametrization become
$c_0=0.01364$, $c_2=0.2731$, $c_4=-0.01545$ and $c_6=0.01975$ with $g_0^2=1.0$
for the plaquette action\cite{sigmaP}, and
$c_0=0.524$, $c_2=0.274$, $c_4=0.105$ and $c_6=0$ with $g_0^2=6/2.4$
for the RG action\cite{sigmaRG}.  We use $\sigma^{1/2}$ = 0.44 GeV
to get $a$ in the Table~\ref{tab:param}.
Gauge fields are updated by the pseudo-heat bath algorithm with five hits,
followed by four over-relaxation sweeps; the combination of these updates is called
an iteration. After 2000 iterations for a thermalization, we calculate
the fermionic correlation functions on the gauge configurations separated by 200 
iterations.
On each $M$ at given $\beta$, different gauge configurations are used to evaluate 
$Z_V$ and $Z_A$,
so that the measurements of  $Z$'s at different $M$ are  independent.
Raw data of $Z_V$(the 3rd and 4th columns) and $Z_A$(the 2nd column) are compiled 
in Table~\ref{tab:resultP} for the plaquette action and
Table~\ref{tab:resultRG} for the RG action.

It has been shown that $Z_V = Z_A$ in perturbation theory for DWQCD\cite{AIKT}, 
and, as already mentioned in the previous section, this equality is well 
satisfied non-perturbatively at $\beta=6.0$ for the plaquette action
and at $\beta = 2.6$ for the RG action.
Although the violation of this equality becomes larger at
stronger coupling(at $\beta=5.8$ for the plaquette action and at $\beta=2.4, 2.2$ 
for the RG action), or at the values of $M$ far away from the ''minimum''
where the chiral symmetry is best realized,
we define $Z_A = Z_V$ for DWQCD in this paper, taking numerical values of 
$Z_V$ as the renormalization factor for both vector and axial-vector currents.
Therefore we discuss $Z_V$ only hereafter.

\subsection{$Z_V$ as a function of $M$}
At each $g^2$, we fit $Z_V$ as a function of $M$ by the formula
\beqa
Z_V^{\rm fit} &=& \frac{B_0 + B_1(M-M_c)+B_2(M-M_c)^2}{1+A_2(M-M_c)^2},
\eeqa
which is suggested by the perturbation theory\cite{AIKT}.
Results for fit parameters $M_C$, $A_2$ and $B_i$($i=0,1,2$) are given
in Tables~\ref{tab:fitM-P} and \ref{tab:fitM-RG}, 
together with $\delta^{\rm max}$, the maximum of the relative errors $\delta$ defined by 
\beqa
\delta &=& (Z_V-Z_V^{\rm fit })/Z_V .
\eeqa
As we observe that $\delta^{\rm max}$ are typically less than 1\% and at most a few \%, 
the fit describes data well. 

\subsection{Finite $a/L$ errors}
Errors of $Z_V$ associated with the lattice spacing are $O((a/L)^2$ in the 
$N_s\rightarrow\infty$ limit, or 
$O(e^{-\alpha N_s}\times a/L)$ at finite $N_s$. 
If $a/L = 1/N_l=1/8$ is kept fixed at all value of $g^2$,  
the scaling violation in $Z_{V,A}$ remains even in the $g^2\rightarrow 0$ limit.
In order to reduce or remove this scaling violation, 
we interpolate or extrapolate $Z_V$ to the fixed value of $L$ at each $g^2$,
using data on two different spatial lattice sizes $L$.
For the interpolation or the extrapolation, we adopt the linear dependence:
\beqa
Z_V(a/L) &=& Z_V + c\frac{a}{L}= Z_V + c\frac{1}{N_l} .
\label{eq:linear}
\eeqa
Since only data at two different $N_l$ are available,
the value of $Z_V$ at fixed $L$ and its error are estimated by
\beqa
Z_V(a/L)&=& \frac{Z_V^2 x_{1}-Z_V^1 x_{2}}{x_{12}} \\
\delta Z_V(a/L)&=& \frac{\vert \delta Z_V^2 x_{1}\vert+\vert \delta Z_V^1 x_{2}\vert}{x_{12}} 
\eeqa
where $Z_V^{i}= Z_V(1/N_i)$, $x_{i}=1/N_i -a/L$ ($i=1,2$) and $x_{12}=1/N_1-1/N_2$
with $N_1=8$ and $N_2 = 4$ or 12, and $\delta$ means the error of the corresponding quantity.
To estimate the systematic uncertainty associated with the assumption eq.~(\ref{eq:linear}),
we alternatively employ the quadratic form:
\beqa
Z_V(a/L) &=& Z_V + c\frac{a^2}{L^2}= Z_V + c\frac{1}{N_l^2} ,
\eeqa
and calculate $Z_V(1/L)$. A difference in $Z_V(a/L)$ between the linear and 
the quadratic dependences is quadratically added to $\delta Z_V(a/L)$ as an estimate 
of the systematic uncertainty, while the central value
of $Z_V(a/L)$ is taken from the value obtained from the linear assumption.

In Tables~\ref{tab:resultP} and \ref{tab:resultRG}, the values of $Z_V(a/L)$ are  
given at $ L=L^* = 8 a(\beta=6.0)$ (the 5th column),
where $a(\beta=6.0)$ is the lattice spacing at $\beta=6.0$ for the plaquette action,
and at $L= \infty$ (the 8th column). 
While the former definition of $Z_V$ contains an $O(a/L^*)$ error,
which vanishes in the continuum limit, the latter one is free from such 
an uncertainty.
By taking the difference of $Z_V$ between $L=L^*$ and $\infty$,
the $a/L^*$ error in $Z_V$ is estimated to be 0.06  at $M=1.8$ and 
$\beta = 6.0$ ($a^{-1}=$2 GeV) for the plaquette action, or
0.02 at $M=1.7$ and $\beta=2.6$ ($a^{-1}=$1.9 GeV) for the RG action.
On the other hand, the error associated with the extrapolation in $L$ is larger 
at $L=\infty$:
0.002 ($L=L^*$) and 0.025 ($L=\infty$) at the previous parameters for the 
plaquette action, and 0.002 and 0.017 for the RG action.
Moreover $Z_V$ at $L=\infty$ monotonically decreases as $M$ increases 
at $a^{-1} < 2$ GeV, while it has the minimum in $M$ at $a^{-1} \ge 2$ GeV.
Only the latter behaviour is observed for $Z_V$ at $L=L^*$.
We suspect that the behaviour of $Z_V$ at $L=\infty$ is related to the existence
of (near) zero eigenvalues for the hermitian Wilson-Dirac operator at $a^{-1} < 2$:
It suggests that the gap of zero eigenvalues for the the hermitian Wilson-Dirac 
operator is closed at $a^{-1} < 2$ for both plaquette and RG actions.
This speculation is consistent with the observation that DWQCD can not 
realize an exact chiral symmetry even in the $N_s\rightarrow\infty$ limit
at $a^{-1} \simeq$ 1 GeV for both gauge actions\cite{chiral},
though the quenched artifact may explain the observation\cite{GS}.

\subsection{$Z_V$ as a function of $M$ and $g^2$}
\begin{figure}[p]
\centering{\includegraphics[width=125mm,angle=0]{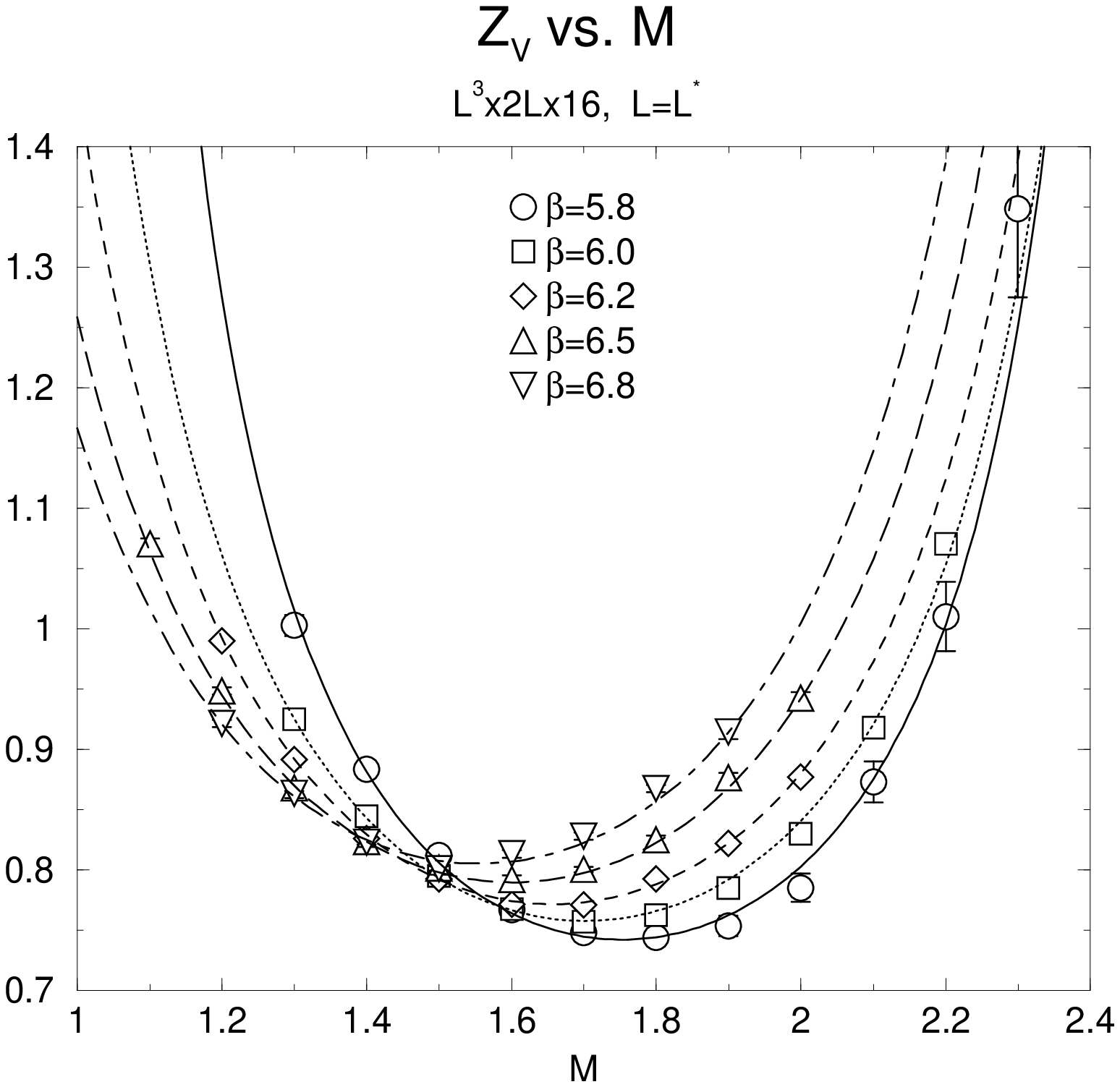}}
\vspace{-0.5cm}
\centering{\includegraphics[width=125mm,angle=0]{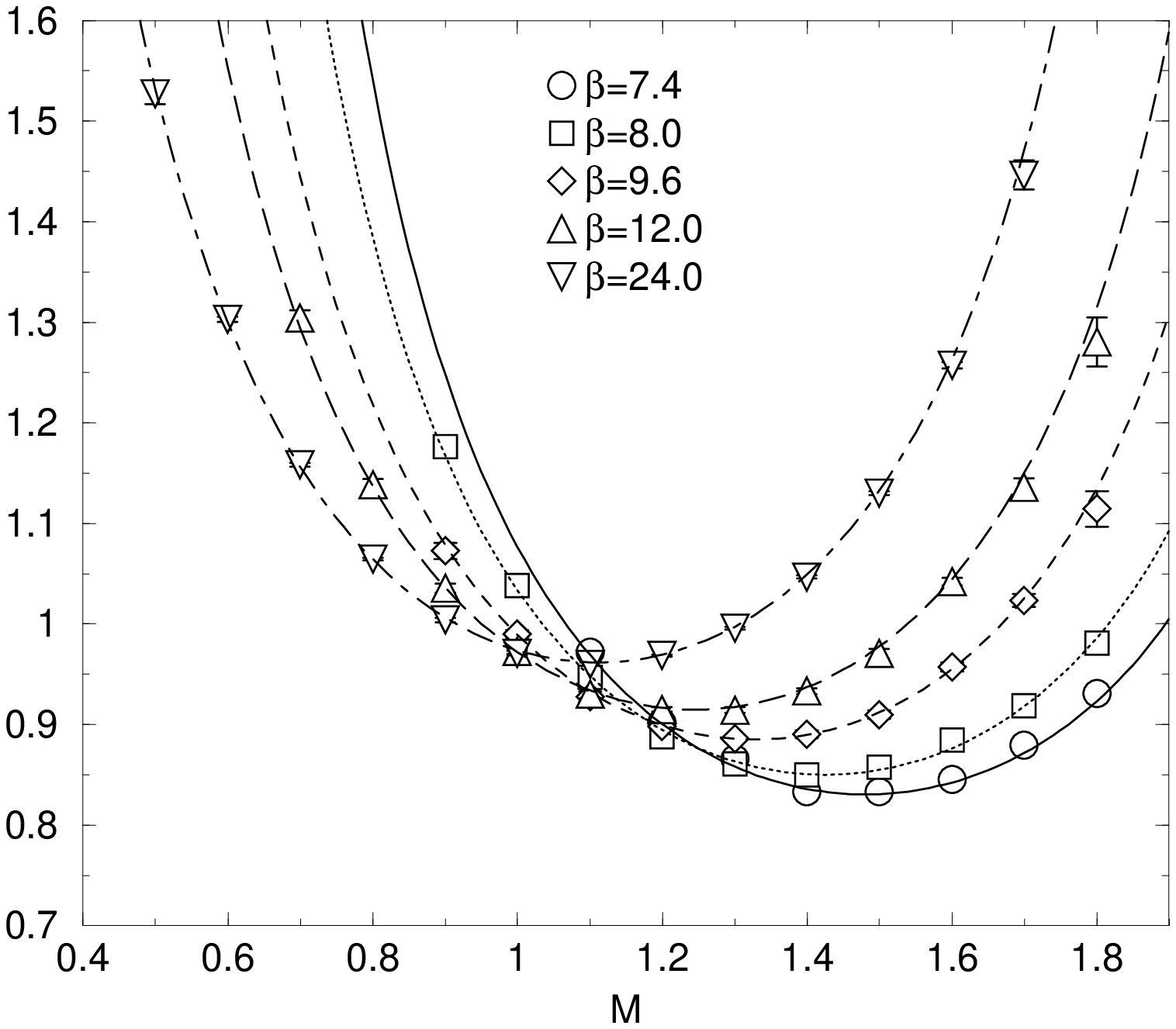}}
\caption{$Z_{V}$ as a function $M$ on $L^3\times 2L\times 16$ with
$L=L^*$ at several values of $\beta$  for the plaquette action.}
\label{fig:fitG-PL}
\end{figure}

\begin{figure}[p]
\centering{\includegraphics[width=125mm,angle=0]{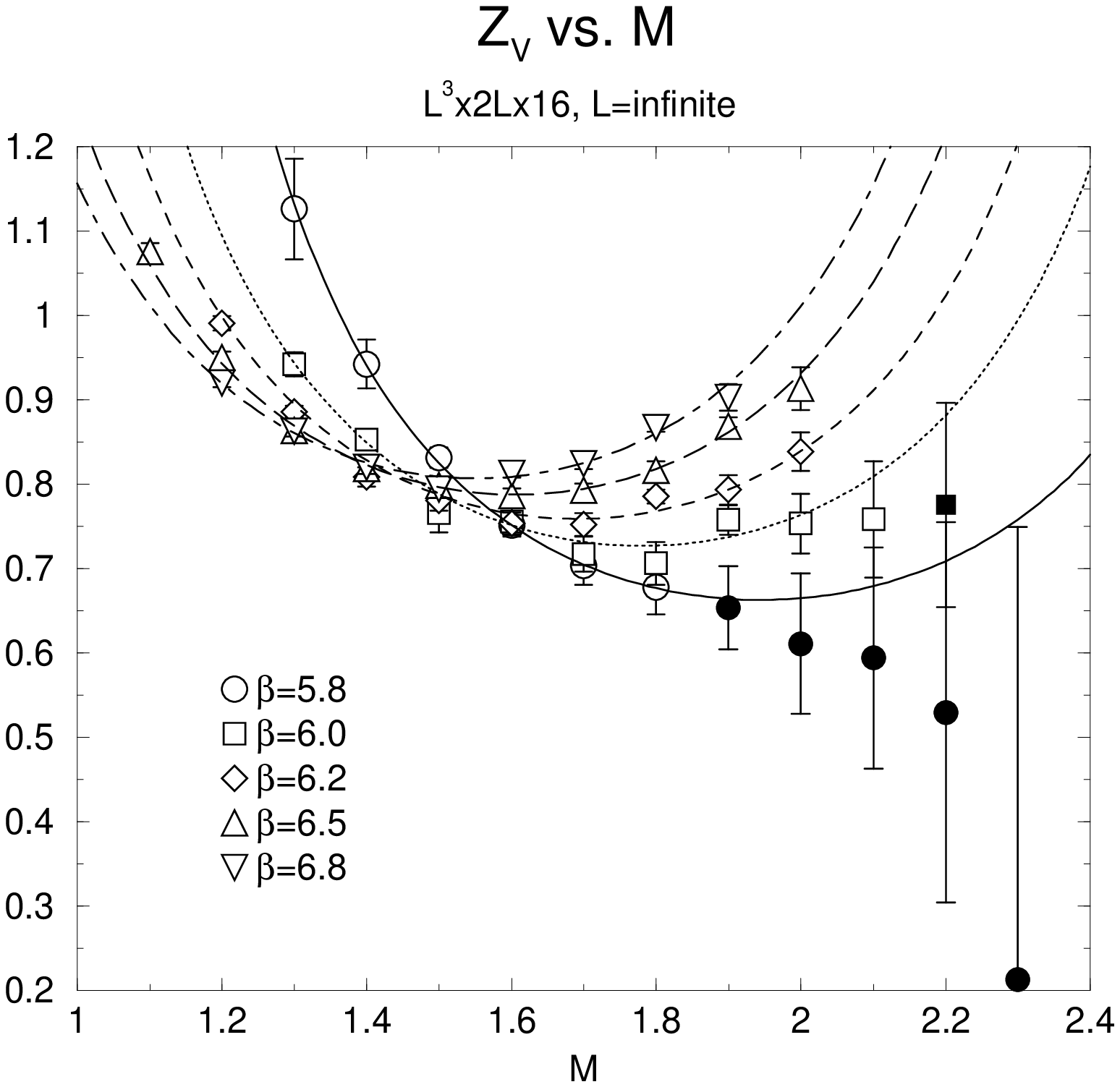}}
\vspace{-0.5cm}
\centering{\includegraphics[width=125mm,angle=0]{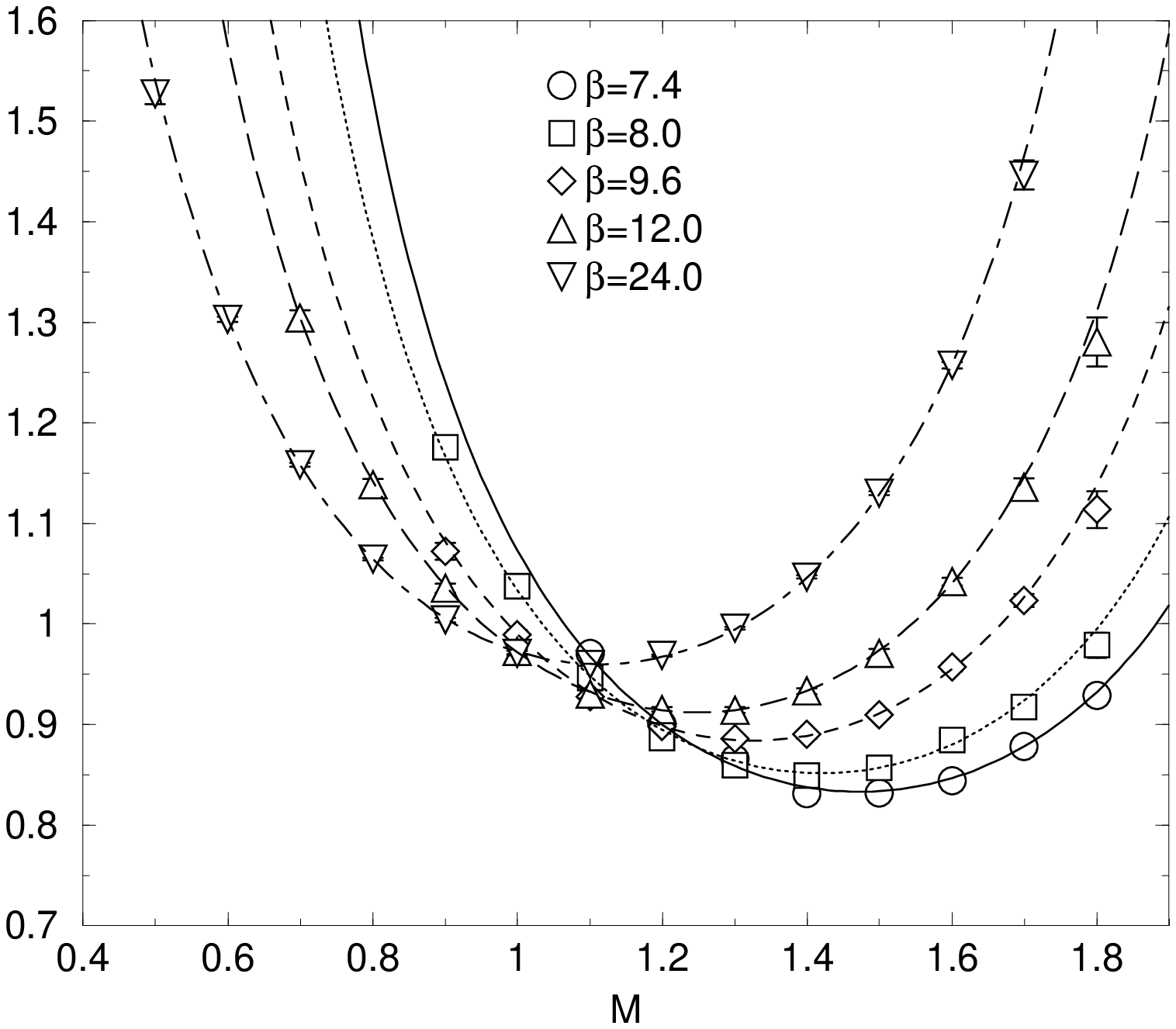}}
\caption{$Z_{V}$ as a function $M$ on $L^3\times 2L\times 16$ with
$L=\infty$ at several values of $\beta$  for the plaquette action.
Solids symbols are excluded for the fits.}
\label{fig:fitG-P}
\end{figure}

\begin{figure}[p]
\centering{\includegraphics[width=125mm,angle=0]{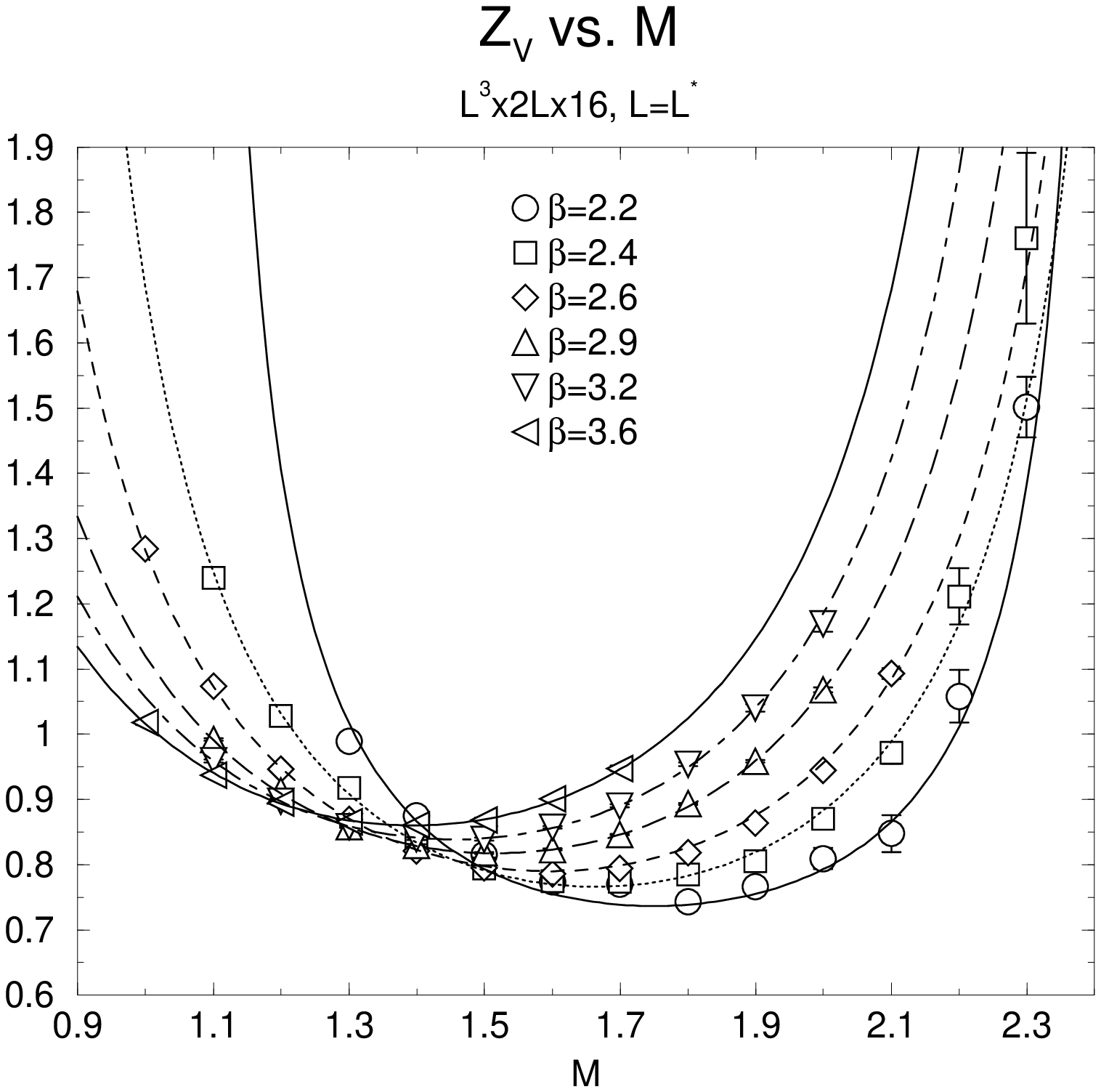}}
\vspace{-0.5cm}
\centering{\includegraphics[width=125mm,angle=0]{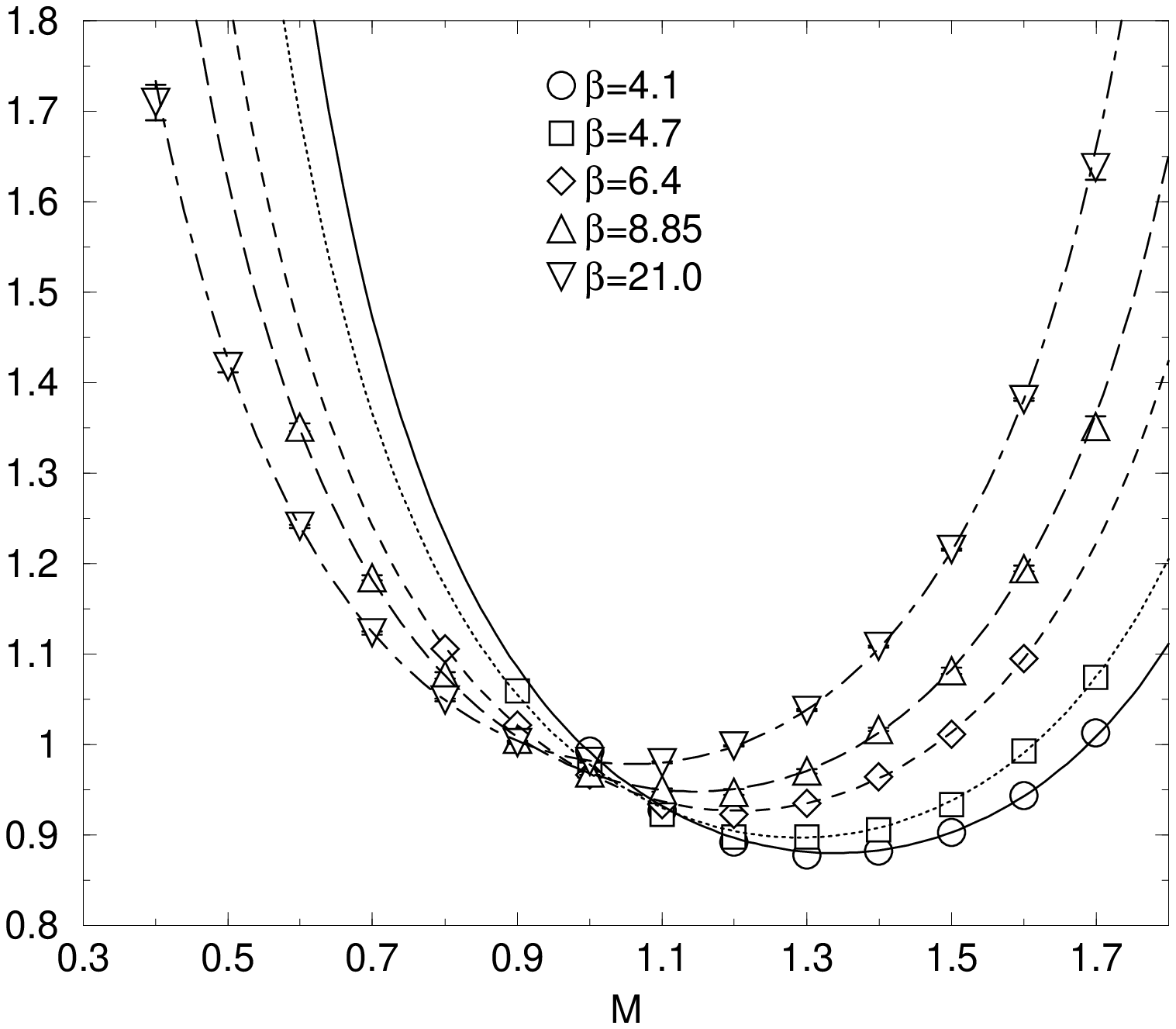}}
\caption{$Z_{V}$ as a function $M$ on $L^3\times 2L\times 16$ with
$L=L^*$ at several values of $\beta$  for the RG action.}
\label{fig:fitG-RGL}
\end{figure}

\begin{figure}[p]
\centering{\includegraphics[width=125mm,angle=0]{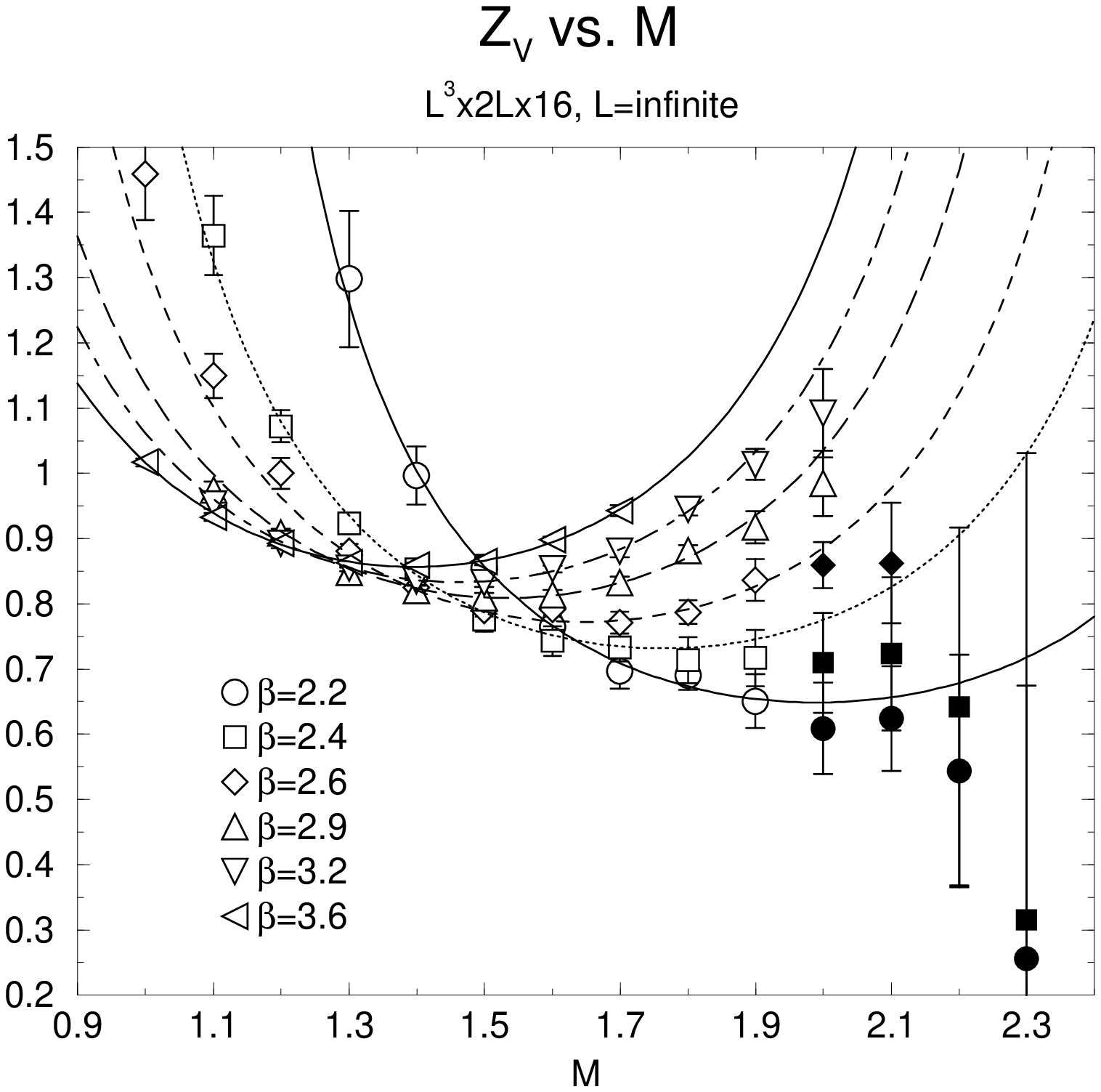}}
\vspace{-0.5cm}
\centering{\includegraphics[width=125mm,angle=0]{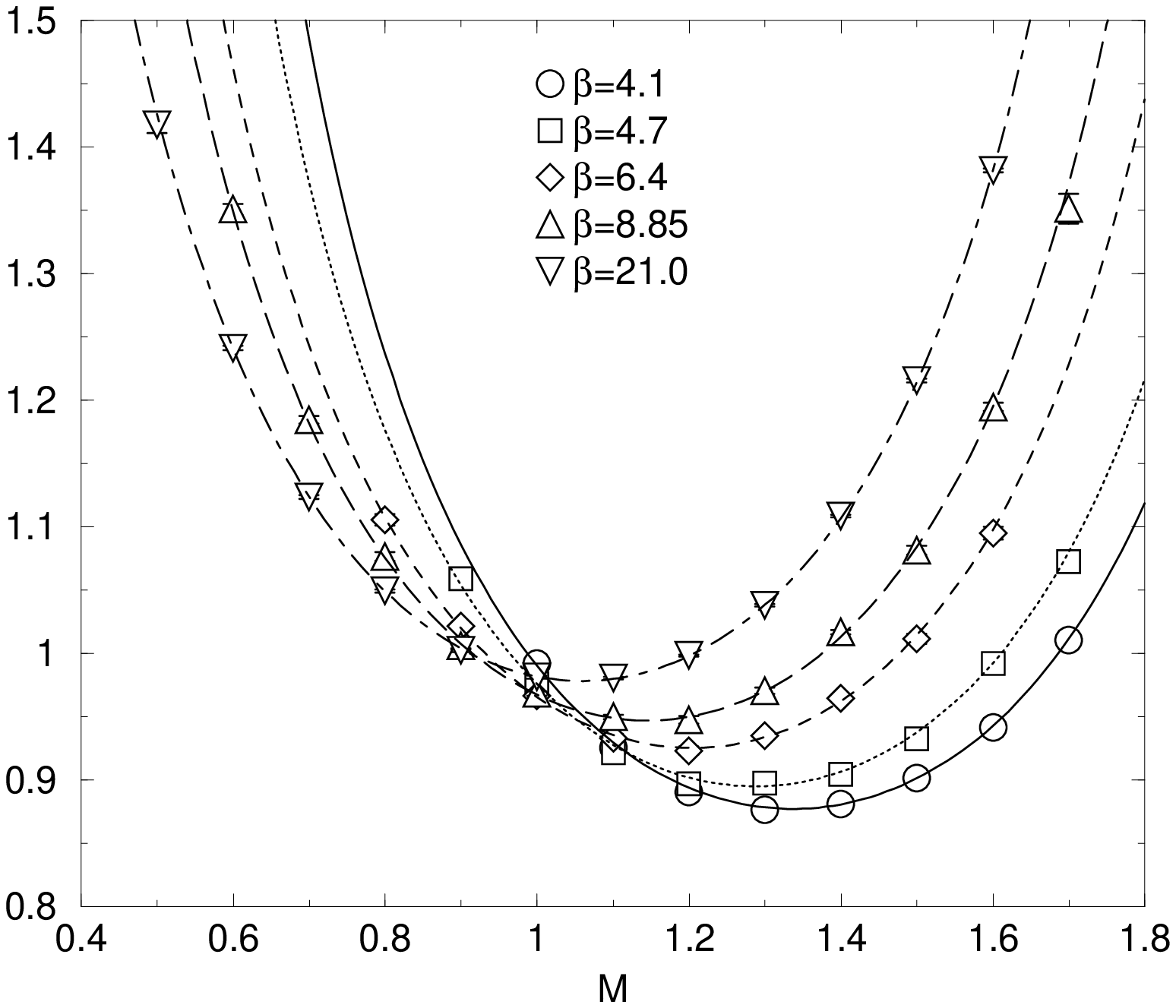}}
\caption{$Z_{V}$ as a function $M$ on $L^3\times 2L\times 16$ with
$L=\infty$ at several values of $\beta$  for the RG action.
Solids symbols are excluded for the fits.}
\label{fig:fitG-RG}
\end{figure}

For the latter uses, we parametrize $Z_V$ as a function of $M$ and $g^2$, in order to
obtain $Z$ at arbitrary (interpolated) values of $\beta$ and $M$.
We adopt the following fitting function suggested by the perturbation theory\cite{AIKT}:
\beqa
Z_V(g^2,M) &=& \frac{B_0(g^2)+B_1(g^2) ( M-M_c(g^2)) +B_2(g^2) ( M-M_c(g^2))^2 }
{1 + A_2(g^2) ( M-M_c(g^2))^2} 
\label{eq:global} \\
M_c(g_2) &=& \dfrac{1+(c_M+a_1) g^2 +a_2 g^4+a_3 g^6}{1+a_1 g^2} \nn \\
B_0(g_2) &=& \dfrac{1+(c_0+a_4) g^2 +a_5 g^4+a_6 g^6}{1+a_4 g^2}\nn \\
A_2(g_2) &=& -\dfrac{1+(d_2+a_7) g^2 +a_9 g^4+a_{10} g^6}{1+a_7 g^2+a_8 g^4}\nn \\
B_1(g_2) &=& c_1 g^2 \nn \\
B_2(g_2) &=& \dfrac{ c_2 g^2}{1+a_{11} g^2} \nn
\eeqa
where $c_M$, $d_2$ and $c_i$($i=0,1,2$) are values of 1-loop coefficients for
$M_c$, $A_2$ and  $B_i$($i=0,1,2$)\cite{AIKT}, which are given by
\beqa
(c_M,d_2,c_0,c_1,c_2)&=&(0.4177, 0.01173,-0.1456, 2.311\times 10^{-3}, 
-5.172\times 10^{-3}) \nn
\eeqa
for the plaquette action and
\beqa
(c_M,d_2,c_0,c_1,c_2)&=&(0.2070, 8.131\times 10^{-3},-0.07449, 
1.912\times 10^{-3}, -4.154\times 10^{-3}) \nn
\eeqa
for the RG action. These constraints make
$M_c$, $A_2$ and $B_i$($i=0,1,2$) consistent with the perturbation theory at 1-loop.
Numerical values of parameters $a_i$ ($i=1\sim 11$)
are given in Table~\ref{tab:fit-g} for both $L=L^*$ and $L=\infty$.
In order to show the quality of the fits, we plot
the fitting curves for $Z_V(a/L^*)$ and $Z_V(0)$  
in Fig.\ref{fig:fitG-PL} and Fig.\ref{fig:fitG-P} for the plaquette action,
and in Fig.\ref{fig:fitG-RGL} and Fig.\ref{fig:fitG-RG}  for the RG action,
respectively.
Furthermore,  we compile $Z_V(g^2,M)$ and the relative deviation
\beqa
\delta_G &=& \dfrac{Z_V-Z_V(g^2,M)}{Z_V}
\eeqa
in the 6th and 7th columns or the 9th and 10th columns of Tables~\ref{tab:resultP} 
and~\ref{tab:resultRG}, where $Z_V = Z_V(a/L^*)$ or $Z_V(0)$, respectively.
In the fit for $Z_V(0)$  we exclude a few points for larger values of $M$ at $\beta=5.8$ and $6.0$ 
for the plaquette action and at $\beta=2.2,2.4,2.6$ for the RG action, 
which are represented by solid symbols in the figures and are 
marked by ''- '' in the tables.
These data for $Z_V(0)$ have large errors and large values of $\delta_G$.
From the figures and tables we observe that the fits works well and 
$\delta_G$ are less than a few \%, 
except a few points at the edges of the range in $M$ employed for the simulations.

\section{Conclusions and Discussions}
\label{sec:concl}

We calculate the renormalization factors for the vector and axial-vector currents in the quenched
DWQCD for both plaquette and RG actions.
After several tests are performed at $a^{-1}\simeq $ 2 GeV,
we obtain $Z_V$, which is assumed to be equal to $Z_A$, at $L=\infty$ as well as $L=L^*$
for the wide ranges of $g^2$ and $M$. We globally fit $Z_V$ as a function of 
$g^2$ and $M$.

We now propose how to use these results for the future simulations of the quenched DWQCD.
\begin{enumerate}
\item For the DWQCD to realize the chiral symmetry well, one should take at least $a^{-1}\ge$ 2 GeV. 
This condition is satisfied at $\beta\ge 6.0$
for the plaquette action or at $\beta \ge 2.6$ for the RG action. 
\item One should choose the optimal value of $M$, which minimizes the violation of the 
chiral symmetry at finite $N_s$. A good candidate is the choice that  $M\simeq M_c(g^2)$, where 
\beqa
M_c(g^2) &=& \frac{ 1 +(c_M + a_1) g^2 + a_2 g^4 + a_3 g^6}{1+a_1 g^2} .
\eeqa
The parameters $a_{1,2,3}$ are given in Table~\ref{tab:fit-g}.
\item If the simulation point at $g^2$ and $M$ can be found in the Table~\ref{tab:resultP}
or Table~\ref{tab:resultRG}, one should use $Z_V$ in the table as the renormalization
factor for the vector or axial-vector current.
To remove $O(a/L)$ errors in $Z_V$, it is better to take $Z_V$ at $L=\infty$,
though the statistical error of $Z_V$ is larger in this case.
One may use $Z_V$ at $L=L^*$ to estimate the size of $O(a/L)$ errors in $Z_V$.
\item If $g^2$ or $M$ for the simulation point is not found in the tables,
one should use the fitting function given in eq.~(\ref{eq:global}) 
with the parameters
in Table~\ref{tab:fit-g}. The error of $Z_V$ is estimated from the errors of $Z_V$
at the nearest points in $g^2$ and $M$, which can be found in the tables.
\end{enumerate}

We are encouraged with the present results to proceed to an extension of the 
present work to scale-dependent cases such as quark masses
and four-quark operators needed for $B_K$, implementing the
SF boundary condition for the domain-wall fermions\cite{AKT}. 

\section*{Acknowledgments}
S.A. thanks Profs. M.~L\"uscher, S.~Sint, P.~Weisz and H.~Wittig
for useful discussions.
This work is supported in part by Grants-in-Aid of the Ministry of Education
(Nos.
12304011, 
12640253, 
12740133, 
13135204 
13640259 
13640260 
14046202 
14740173
15204015 
15540251 
15540279
15740134 
).

\clearpage
\begin{longtable}{|ll|c|c|ll|c|c|}
\caption{Simulation parameters}
\label{tab:param} \\
\hline
\multicolumn{4}{|c|}{Plaquette}&\multicolumn{4}{c|}{RG improved}\\
\hline
$\beta$ & $a^{-1}$(GeV)&\# of conf. &\# of conf. &
$\beta$ & $a^{-1}$(GeV)&\# of conf. &\# of conf. \\
\hline
\endfirsthead
& & {$8^3\times 16\times 16$}
& {$4^3\times 8\times 16$}
& & & {$8^3\times 16\times 16$} 
&{$4^3\times 8\times 16$}\\
\hline
    &     &     &     & 2.2 & 1.0 & 100 &  100-200 \\
5.8 & 1.4 & 100 & 100 & 2.4 & 1.4 & 100 &  100-200\\
\hline
& &{$8^3\times 16\times 16$}
& {$12^3\times 24\times 16$} 
& & & {$8^3\times 16\times 16$}
& {$12^3\times 24\times 16$}\\
\hline
6.0 & 2.0 & 100 & 20   & 2.6 & 1.9 & 100 &  30\\
6.2 & 2.7 & 100 & 10-30& 2.9 & 2.9 & 100 &15-50 \\
6.5 & 4.1 & 100 & 15-25& 3.2 & 4.3 & 100 &20-25 \\
6.8 & 6.1 & 100 & 10-15& 3.6 & 6.8 & 100 &20-25 \\
7.4 & 12  & 100 & 10-20& 4.1 & 12  & 100 &20-25 \\
8.0 & 25  &  40 & 7-15 & 4.7 & 23  &  40 &10-20 \\
9.6 & 156 &  40 &  10  & 6.4 & 154 &  40 &10-20 \\
12.0& 2502&  20 &  10  & 8.85& 2523&  20 &10-15 \\
24.0&3.2$\times 10^9$ &20& 10 & 21.0 & 3.6$\times 10^9$ & 20 &10\\
\hline
\end{longtable}

\pagebreak
\begin{longtable}{|c|c|cc|ccc|ccc|}
\caption{Results for the plaquette action. }
\label{tab:resultP}\\
\hline
& $Z_A$ &
\multicolumn{2}{c|}{$Z_V$}&
\multicolumn{3}{c|}{$Z_V$ at $L=L^*$}&
\multicolumn{3}{c|}{$Z_V$ at $L=\infty$}\\
\hline
$M$ & $8^3\times 16$ & $8^3\times 16$ & 
$4^3\times 8$& 
$Z_V^L$ & Fit & $\delta_G$(\%) &
$Z_V^L$ & Fit & $\delta_G$(\%) \\
\hline
\endfirsthead
\caption[]{(continued)}\\
\hline
& $Z_A$ &
\multicolumn{2}{c|}{$Z_V$}&
\multicolumn{3}{c|}{$Z_V$ at $L=L^*$}&
\multicolumn{3}{c|}{$Z_V$ at $L=\infty$}\\
\hline
$M$ & $8^3\times 16$ & $8^3\times 16$ & 
$12^3\times 24$& 
$Z_V^L$ & Fit & $\delta_G$(\%) &
$Z_V^L$ & Fit & $\delta_G$(\%) \\
\hline
\endhead
\multicolumn{10}{|l|}{$\beta=5.8$, $a^{-1}$=1.4 GeV, $L^*/a=5.6$}\\
\hline
1.3& 0.7724(119)& 1.0398(69)& 0.9535(78)&1.0028(87)  &1.0153 &1.2 &1.1260(596) &1.1318 &0.5 \\
1.4& 0.7203(94) & 0.9013(37)& 0.8603(64)&0.8837(48)  &0.8815 &0.2 &0.9423(290) &0.9406 &0.2 \\
1.5& 0.6914(104)& 0.8179(28)& 0.7852(63)&0.8122(33)  &0.8062 &0.7 &0.8313(123) &0.8250 &0.8 \\
1.6& 0.6603(83) & 0.7624(41)& 0.7731(90)&0.7670(46)  &0.7641 &0.4 &0.7516(141) &0.7514 &0.02\\
1.7& 0.6383(98) & 0.7348(30)& 0.7659(68)&0.7481(42)  &0.7449 &0.4 &0.7036(226) &0.7046 &0.1 \\
1.8& 0.6377(82) & 0.7245(30)& 0.7711(41)&0.7444(45)  &0.7443 &0.01&0.6778(319) &0.6768 &0.1 \\
1.9& 0.6309(79) & 0.7238(42)& 0.7939(134)&0.7538(85) &0.7624 &1.1 &0.6537(493) &0.6641 &1.6 \\
2.0& 0.6290(84) & 0.7330(55)& 0.8550(129)&0.7853(118)&0.8029 &2.2 &0.6110(831) & - & - \\
2.1& 0.6710(115)& 0.7894(54)& 0.9848(122)&0.8732(171)&0.8755 &0.3 &0.5940(1313)& - & - \\
2.2& 0.6722(149)& 0.8660(81)& 1.2024(160)&1.0102(287)&1.0037 &0.6 &0.5296(2254)& - & - \\
2.3& 0.7080(203)&1.0079(106)& 1.8025(796)&1.3484(736)&1.2499 &7.3 &0.2132(5361)& - & - \\
\hline
\multicolumn{10}{|l|}{$\beta=6.0$, $a^{-1}$=2.0 GeV, $L^*/a=8.0$}\\
\hline
1.3& 0.9002(21)& 0.9252(30)& 0.9308(36)&0.9252(30)&0.9253 &0.01 &0.9421(141)&0.9438 &0.2 \\
1.4& 0.8330(21)& 0.8446(25)& 0.8472(32)&0.8446(25)&0.8431 &0.2 &0.8526(112) &0.8491 &0.4 \\
1.5& 0.7884(20)& 0.7950(22)& 0.7927(44)&0.7950(22)&0.7935 &0.2 &0.7656(227) &0.7885 &3.0 \\
1.6& 0.7633(19)& 0.7674(21)& 0.7635(50)&0.7674(21)&0.7666 &0.1 &0.7555(163) &0.7511 &0.6 \\
1.7& 0.7541(19)& 0.7572(21)& 0.7438(42)&0.7572(21)&0.7578 &0.1 &0.7169(208) &0.7316 &2.0 \\
1.8& 0.7589(20)& 0.7628(21)& 0.7439(35)&0.7628(21)&0.7659 &0.4 &0.7062(254) &0.7272 &3.0 \\
1.9& 0.7781(23)& 0.7852(21)& 0.7761(43)&0.7852(21)&0.7921 &0.9 &0.7577(175) &0.7375 &3.0 \\
2.0& 0.8153(30)& 0.8300(25)& 0.8044(58)&0.8300(25)&0.8406 &1.3 &0.7532(356) &0.7636 &1.4 \\
2.1& 0.8802(52)& 0.9179(34)& 0.8647(84)&0.9179(34)&0.9210 &0.3 &0.7583(690) &0.8092 &6.7 \\
2.2& 0.9941(95)& 1.0706(48)& 0.9722(81)&1.0706(48)&1.0535 &1.6 &0.7754(1209)& - & - \\
\hline
\pagebreak
\multicolumn{10}{|l|}{$\beta=6.2$, $a^{-1}$=2.7 GeV, $L^*/a=10.8$}\\
\hline
1.2& 0.9734(15)& 0.9896(21)& 0.9899(25)&0.9898(20)&0.9926 &0.3 & 0.9904(87)&0.9982 &0.8 \\
1.3& 0.8858(11)& 0.8937(14)& 0.8911(19)&0.8917(15)&0.8921 &0.04& 0.8859(70)&0.8947 &1.0 \\ 
1.4& 0.8294(07)& 0.8321(13)& 0.8242(18)&0.8260(15)&0.8295 &0.4 &0.8084(113)&0.8284 &2.5 \\
1.5& 0.7957(10)& 0.7969(16)& 0.7915(32)&0.7927(25)&0.7923 &0.1 &0.7805(121)&0.7872 &0.9 \\
1.6& 0.7764(09)& 0.7781(13)& 0.7700(27)&0.7718(21)&0.7744 &0.3 &0.7537(129)&0.7651 &1.5 \\
1.7& 0.7766(10)& 0.7778(14)& 0.7693(28)&0.7712(23)&0.7731 &0.2 &0.7524(135)&0.7589 &0.9 \\
1.8& 0.7892(11)& 0.7949(18)& 0.7918(21)&0.7925(17)&0.7884 &0.5 &0.7855(82) &0.7681 &2.2 \\
1.9& 0.8196(14)& 0.8319(19)& 0.8190(23)&0.8219(19)&0.8223 &0.05&0.7932(174)&0.7936 &0.05\\
2.0& 0.8659(27)& 0.8908(22)& 0.8734(30)&0.8772(24)&0.8802 &0.3 &0.8386(231)&0.8389 &0.04\\
\hline
\multicolumn{10}{|l|}{$\beta=6.5$, $a^{-1}$=4.1 GeV, $L^*/a=16.4$}\\
\hline
1.1& 1.0458(13)& 1.0656(18)& 1.0689(26)&1.0707(42)&1.0635 &0.7 &1.0757(95) &1.0588 &1.6 \\
1.2& 0.9416(10)& 0.9469(16)& 0.9480(19)&0.9485(31)&0.9449 &0.4 &0.9500(67) &0.9431 &0.7 \\
1.3& 0.8715(08)& 0.8728(13)& 0.8698(16)&0.8682(26)&0.8700 &0.2 &0.8638(65) &0.8690 &0.6 \\
1.4& 0.8285(08)& 0.8290(12)& 0.8256(16)&0.8238(27)&0.8236 &0.02&0.8188(68) &0.8226 &0.5 \\
1.5& 0.8042(06)& 0.8042(11)& 0.8024(11)&0.8014(19)&0.7981 &0.4 &0.7987(47) &0.7966 &0.3 \\
1.6& 0.7963(07)& 0.7976(08)& 0.7939(23)&0.7919(36)&0.7898 &0.3 &0.7864(84) &0.7876 &0.2 \\
1.7& 0.8051(07)& 0.8082(12)& 0.8031(14)&0.8004(23)&0.7976 &0.3 &0.7930(77) &0.7945 &0.2 \\
1.8& 0.8246(10)& 0.8335(14)& 0.8279(23)&0.8249(38)&0.8226 &0.3 &0.8167(101)&0.8181 &0.2 \\
1.9& 0.8687(17)& 0.8853(17)& 0.8797(26)&0.8767(41)&0.8682 &1.0 &0.8685(108)&0.8615 &0.8 \\
2.0& 0.9345(29)& 0.9730(21)& 0.9531(25)&0.9424(52)&0.9419 &0.1 &0.9132(254)&0.9311 &2.0 \\
\hline
\pagebreak
\multicolumn{10}{|l|}{$\beta=6.8$, $a^{-1}$=6.1 GeV, $L^*/a=24.4$}\\
\hline
1.2& 0.9250(07)& 0.9250(13)& 0.9233(13)&0.9215(31)&0.9211 &0.03&0.9198(52) &0.9192 &0.1 \\
1.3& 0.8667(07)& 0.8656(12)& 0.8647(19)&0.8638(41)&0.8610 &0.3 &0.8629(64) &0.8605 &0.3 \\
1.4& 0.8317(06)& 0.8304(10)& 0.8269(20)&0.8232(43)&0.8250 &0.2 &0.8197(75) &0.8253 &0.7 \\
1.5& 0.8156(06)& 0.8157(09)& 0.8085(19)&0.8011(49)&0.8078 &0.8 &0.7940(105)&0.8088 &1.9 \\
1.6& 0.8140(07)& 0.8147(12)& 0.8141(14)&0.8134(31)&0.8070 &0.8 &0.8129(49) &0.8088 &0.5 \\
1.7& 0.8300(07)& 0.8359(10)& 0.8319(11)&0.8279(30)&0.8227 &0.6 &0.8239(63) &0.8253 &0.2 \\
1.8& 0.8575(11)& 0.8692(14)& 0.8685(13)&0.8677(31)&0.8568 &1.3 &0.8670(50) &0.8603 &0.8 \\
1.9& 0.9116(19)& 0.9393(16)& 0.9271(17)&0.9147(63)&0.9142 &0.1 &0.9027(158)&0.9189 &1.8 \\
\hline
\multicolumn{10}{|l|}{$\beta=7.4$, $a^{-1}$=12 GeV, $L^*/a=48$}\\
\hline
1.1& 0.9761(06)& 0.9762(11)& 0.9744(14)&0.9717(41)&0.9693 &0.2 &0.9708(53)&0.9677 &0.3 \\
1.2& 0.9090(07)& 0.9078(09)& 0.9054(12)&0.9019(37)&0.9005 &0.2 &0.9007(48)&0.9002 &0.1 \\
1.3& 0.8646(05)& 0.8631(08)& 0.8642(14)&0.8658(37)&0.8582 &0.9 &0.8663(45)&0.8588 &0.9 \\
1.4& 0.8437(05)& 0.8441(08)& 0.8398(16)&0.8335(53)&0.8359 &0.3 &0.8313(72)&0.8376 &0.8 \\
1.5& 0.8394(05)& 0.8403(08)& 0.8374(13)&0.8332(41)&0.8309 &0.3 &0.8318(54)&0.8337 &0.2 \\
1.6& 0.8497(05)& 0.8535(08)& 0.8504(23)&0.8457(63)&0.8424 &0.4 &0.8442(80)&0.8467 &0.3 \\
1.7& 0.8748(07)& 0.8854(09)& 0.8829(20)&0.8792(55)&0.8718 &0.8 &0.8780(69)&0.8782 &0.02\\
1.8& 0.9150(14)& 0.9418(11)& 0.9375(10)&0.9311(43)&0.9232 &0.8 &0.9289(63)&0.9325 &0.4 \\
\hline
\pagebreak
\multicolumn{10}{|l|}{$\beta=8.0$, $a^{-1}$=25 GeV, $L^*/a=100$}\\
\hline
0.9& 1.1644(16)& 1.1842(22)& 1.1814(20)&1.1764(73) &1.1674 &0.8 &1.1757(82) &1.1669 &0.8 \\
1.0& 1.0421(08)& 1.0437(13)& 1.0431(14)&1.0383(49) &1.0340 &0.4 &1.0378(55) &1.0337 &0.4 \\
1.1& 0.9578(08)& 0.9575(12)& 0.9538(10)&0.9471(51) &0.9486 &0.2 &0.9462(60) &0.9485 &0.2 \\
1.2& 0.9031(07)& 0.9024(09)& 0.8972(17)&0.8881(71) &0.8946 &0.7 &0.8868(83) &0.8948 &0.9 \\
1.3& 0.8715(05)& 0.8712(10)& 0.8674(11)&0.8607(51) &0.8634 &0.3 &0.8598(60) &0.8640 &0.5 \\
1.4& 0.8576(06)& 0.8565(09)& 0.8542(09)&0.8501(37) &0.8506 &0.1 &0.8495(43) &0.8520 &0.3 \\
1.5& 0.8601(06)& 0.8615(11)& 0.8600(07)&0.8575(32) &0.8548 &0.3 &0.8572(36) &0.8572 &0.0 \\
1.6& 0.8790(08)& 0.8846(14)& 0.8846(11)&0.8847(40) &0.8763 &0.9 &0.8847(44) &0.8802 &0.5 \\
1.7& 0.9164(11)& 0.9341(10)& 0.9284(19)&0.9186(78) &0.9180 &0.1 &0.9172(91) &0.9241 &0.8 \\
1.8& 0.9738(21)& 1.0074(18)& 0.9979(14)&0.9812(105)&0.9860 &0.5 &0.9789(127)&0.9953 &1.7 \\
\hline                                   
\multicolumn{10}{|l|}{$\beta=9.6$, $a^{-1}$=156 GeV, $L^*/a=624$}\\
\hline
0.9& 1.0897(07)& 1.0921(11)& 1.0855(08)&1.0726(83) &1.0787 &0.6 &1.0723(86)&1.0817 &0.9 \\
1.0& 0.9994(06)& 0.9973(09)& 0.9947(09)&0.9897(43) &0.9892 &0.1 &0.9896(44)&0.9901 &0.1 \\
1.1& 0.9412(05)& 0.9395(07)& 0.9356(07)&0.9279(52) &0.9324 &0.5 &0.9278(54)&0.9322 &0.5 \\
1.2& 0.9053(05)& 0.9032(09)& 0.9016(05)&0.8984(30) &0.8995 &0.1 &0.8983(31)&0.8986 &0.03\\
1.3& 0.8895(05)& 0.8884(07)& 0.8876(06)&0.8859(25) &0.8858 &0.01&0.8859(25)&0.8847 &0.1 \\
1.4& 0.8928(06)& 0.8938(10)& 0.8926(11)&0.8903(40) &0.8899 &0.05&0.8903(41)&0.8887 &0.2 \\
1.5& 0.9129(04)& 0.9181(07)& 0.9154(08)&0.9099(41) &0.9120 &0.2 &0.9098(42)&0.9112 &0.2 \\
1.6& 0.9471(08)& 0.9621(09)& 0.9605(12)&0.9573(44) &0.9551 &0.2 &0.9572(44)&0.9550 &0.2 \\
1.7& 1.0023(17)& 1.0362(12)& 1.0319(10)&1.0236(63) &1.0253 &0.2 &1.0235(64)&1.0266 &0.3 \\
1.8& 1.0946(32)& 1.1568(17)& 1.1425(18)&1.1145(178)&1.1346 &1.8&1.1139(183)&1.1383 &2.2 \\
\hline
\pagebreak
\multicolumn{10}{|l|}{$\beta=12.0$, $a^{-1}$=2502 GeV, $L^*/a=10008$}\\
\hline
0.7& 1.2733(14)& 1.3174(18)& 1.3131(14)&1.3043(76)&1.2946 &0.7 &1.3043(76)&1.3051 &0.1 \\
0.8& 1.1437(07)& 1.1500(12)& 1.1461(09)&1.1383(59)&1.1370 &0.1 &1.1383(60)&1.1418 &0.3 \\
0.9& 1.0467(06)& 1.0440(10)& 1.0411(06)&1.0356(44)&1.0364 &0.1 &1.0355(44)&1.0381 &0.2 \\
1.0& 0.9803(06)& 0.9759(10)& 0.9747(04)&0.9724(26)&0.9723 &0.01&0.9724(26)&0.9720 &0.04\\
1.1& 0.9405(04)& 0.9369(06)& 0.9346(09)&0.9301(39)&0.9341 &0.4 &0.9301(39)&0.9326 &0.3 \\
1.2& 0.9200(03)& 0.9179(05)& 0.9170(05)&0.9151(22)&0.9166 &0.2 &0.9151(22)&0.9142 &0.1 \\
1.3& 0.9202(06)& 0.9199(07)& 0.9181(05)&0.9145(31)&0.9175 &0.3 &0.9145(31)&0.9145 &0.0 \\
1.4& 0.9360(05)& 0.9391(07)& 0.9372(03)&0.9332(29)&0.9370 &0.4 &0.9332(29)&0.9336 &0.04\\
1.5& 0.9685(06)& 0.9806(09)& 0.9772(08)&0.9705(50)&0.9776 &0.7 &0.9705(50)&0.9738 &0.3 \\
1.6& 1.0200(14)& 1.0473(14)& 1.0456(03)&1.0423(39)&1.0448 &0.2 &1.0423(39)&1.0408 &0.1 \\
1.7& 1.1011(30)& 1.1572(12)& 1.1501(08)&1.1360(91)&1.1500 &1.2 &1.1359(92)&1.1460 &0.9 \\
1.8& 1.2299(54)& 1.3400(20)&1.3202(15)&1.2805(245)&1.3150 &2.7&1.2805(246)&1.3114 &2.4 \\
\hline                                   
\multicolumn{10}{|l|}{$\beta=24.0$, $a^{-1}=3.2\times 10^9$ GeV, $L^*/a=
1.28\times 10^{10}$}\\
\hline                                   
0.5& 1.4168(27)& 1.5526(11)& 1.5441(05)&1.5272(106)&1.5323 &0.3 &1.5272(106)&1.5406 &0.9 \\
0.6& 1.2735(09)& 1.3079(07)& 1.3063(04)&1.3030(27) &1.3003 &0.2 &1.3030(27) &1.3043 &0.1 \\
0.7& 1.1603(02)& 1.1604(05)& 1.1597(06)&1.1583(21) &1.1565 &0.2 &1.1583(21) &1.1582 &0.01\\
0.8& 1.0746(03)& 1.0669(04)& 1.0662(03)&1.0647(14) &1.0644 &0.03&1.0647(14) &1.0647 &0.0 \\
0.9& 1.0154(03)& 1.0083(05)& 1.0069(04)&1.0040(22) &1.0064 &0.2 &1.0040(22) &1.0059 &0.2 \\
1.0& 0.9802(03)& 0.9752(04)& 0.9738(03)&0.9710(20) &0.9738 &0.3 &0.9710(20) &0.9725 &0.2 \\
1.1& 0.9666(02)& 0.9629(03)& 0.9619(02)&0.9600(15) &0.9619 &0.2 &0.9600(15) &0.9602 &0.02\\
1.2& 0.9716(02)& 0.9696(03)& 0.9693(02)&0.9685(09) &0.9695 &0.1 &0.9685(09) &0.9673 &0.1 \\
1.3& 0.9946(03)& 0.9976(04)& 0.9969(02)&0.9956(13) &0.9974 &0.2 &0.9956(13) &0.9947 &0.1 \\
1.4& 1.0350(03)& 1.0493(06)& 1.0485(02)&1.0468(17) &1.0492 &0.2 &1.0468(17) &1.0461 &0.1 \\
1.5& 1.0959(08)& 1.1324(06)& 1.1316(04)&1.1300(19) &1.1328 &0.2 &1.1300(19) &1.1290 &0.1 \\
1.6& 1.1820(23)& 1.2638(06)& 1.2615(04)&1.2571(32) &1.2630 &0.5 &1.2571(32) &1.2584 &0.1 \\
1.7& 1.3151(40)& 1.4815(07)& 1.4698(08)&1.4464(143)&1.4706 &1.7 &1.4464(143)&1.4647 &1.3 \\
\hline
\end{longtable}

\begin{longtable}{|c|c|cc|ccc|ccc|}
\caption{Results for the RG action. }
\label{tab:resultRG}\\
\hline
& $Z_A$ &
\multicolumn{2}{c|}{$Z_V$}&
\multicolumn{3}{c|}{$Z_V$ at $L=L^*$}&
\multicolumn{3}{c|}{$Z_V$ at $L=\infty$}\\
\hline
$M$ & $8^3\times 16$ & $8^3\times 16$ & 
$4^3\times 8$& 
$Z_V^L$ & Fit & $\delta_G$(\%) &
$Z_V^L$ & Fit & $\delta_G$(\%) \\
\hline
\endfirsthead
\caption[]{(continued)}\\
\hline
& $Z_A$ &
\multicolumn{2}{c|}{$Z_V$}&
\multicolumn{3}{c|}{$Z_V$ at $L=L^*$}&
\multicolumn{3}{c|}{$Z_V$ at $L=\infty$}\\
\hline
$M$ & $8^3\times 16$ & $8^3\times 16$ & 
$12^3\times 24$& 
$Z_V^L$ & Fit & $\delta_G$(\%) &
$Z_V^L$ & Fit & $\delta_G$(\%) \\
\hline
\endhead
\multicolumn{10}{|l|}{$\beta=2.2$, $a^{-1}$=1.0 GeV, $L^*/a=4.0$}\\
\hline
1.3& 0.4365(102)& 1.1434(62)&0.9888(121)&0.9888(121)&1.0170 &2.9 &1.2980(1045)&1.2608 &2.9\\
1.4& 0.4588(109)& 0.9356(55)&0.8746(136)&0.8746(136)&0.8671 &0.9 &0.9967(443) &1.0040 &0.7\\
1.5& 0.4438(95) & 0.8356(51)&0.8160(104)&0.8160(104)&0.7934 &2.8 &0.8552(196) &0.8575 &0.3\\
1.6& 0.4489(108)& 0.7693(59)&0.7741( 89)&0.7741( 89)&0.7553 &2.4 &0.7646(151) &0.7667 &0.3\\
1.7& 0.4429(95) & 0.7336(41)&0.7701( 81)&0.7701( 81)&0.7388 &4.1 &0.6970(270) &0.7089 &1.7\\
1.8& 0.4661(181)& 0.7165(51)&0.7429( 93)&0.7429( 93)&0.7388 &0.6 &0.6901(223) &0.6732 &2.4\\
1.9& 0.4580(109)& 0.7086(49)&0.7664(119)&0.7664(119)&0.7553 &1.4 &0.6507(415) &0.6540 &0.5\\
2.0& 0.4615(130)& 0.7092(65)&0.8094(162)&0.8094(162)&0.7931 &2.0 &0.6090(700) & - & - \\
2.1& 0.4516(130)& 0.7360(55)&0.8478(282)&0.8478(282)&0.8657 &2.1 &0.6243(804) & - & - \\
2.2& 0.4533(130)& 0.8019(12)&1.0583(401)&1.0583(401)&1.0115 &4.4 &0.5437(1778)& - & - \\
2.3& 0.4641(132)& 0.8787(14)&1.5012(463)&1.5012(463)&1.3809 &8.0 &0.2562(4185)& - & - \\
\hline
\multicolumn{10}{|l|}{$\beta=2.4$, $a^{-1}$=1.4 GeV, $L^*/a=5.6$}\\
\hline
1.1& 0.9735(137)& 1.2773(65)& 1.1902(116)&1.2400(94)  &1.2486 &0.7 &1.3644(606) &1.32455 &2.9 \\
1.2& 0.9350(90) & 1.0415(57)& 1.0109( 84)&1.0284(55)  &1.0319 &0.3 &1.0721(248) &1.07925 &0.7 \\
1.3& 0.8645(66) & 0.9193(43)& 0.9151( 85)&0.9175(44)  &0.9084 &1.0 &0.9235(124) &0.93494 &1.2 \\
1.4& 0.8143(52) & 0.8438(41)& 0.8348( 78)&0.8399(41)  &0.8343 &0.7 &0.8527(128) &0.84436 &1.0 \\
1.5& 0.7680(37) & 0.7887(40)& 0.8020(141)&0.7944(65)  &0.7911 &0.4 &0.7753(185) &0.78675 &1.5 \\
1.6& 0.7508(44) & 0.7654(39)& 0.7877(157)&0.7749(73)  &0.7700 &0.6 &0.7430(230) &0.75191 &1.2 \\
1.7& 0.7441(39) & 0.7615(42)& 0.7898( 50)&0.7736(39)  &0.7675 &0.8 &0.7333(212) &0.73465 &0.2 \\
1.8& 0.7546(51) & 0.7636(37)& 0.8137( 92)&0.7851(61)  &0.7832 &0.2 &0.7135(354) &0.73266 &2.7 \\
1.9& 0.7613(52) & 0.7787(36)& 0.8405(119)&0.8052(74)  &0.8195 &1.8 &0.7170(434) &0.74565 &4.0 \\
2.0& 0.7859(90) & 0.8219(45)& 0.9342(130)&0.8701(110) &0.8832 &1.5 &0.7096(765) & - & - \\
2.1& 0.8461(143)& 0.8969(49)& 1.0704(175)&0.9712(163) &0.9888 &1.8 &0.7234(1174)& - & - \\
2.2& 0.8927(238)&1.0406(116)& 1.4389(650)&1.2113(433) &1.1692 &3.5 &0.6423(2744)& - & - \\
2.3& 0.9352(487)&1.3270(183)&2.3382(2358)&1.7604(1309)&1.5119 &14.1&0.3157(7151)& - & - \\
\hline
\pagebreak
\multicolumn{10}{|l|}{$\beta=2.6$, $a^{-1}$=1.9 GeV, $L^*/a=7.6$}\\
\hline
1.0& 1.1837(49)& 1.2926(41)& 1.3479(75)&1.2839(53)&1.2828 &0.1 &1.4586(706)&1.3310 &8.7 \\
1.1& 1.0498(23)& 1.0772(27)& 1.1013(55)&1.0734(33)&1.0729 &0.04&1.1495(338)&1.1012 &4.2 \\
1.2& 0.9440(15)& 0.9496(19)& 0.9663(41)&0.9469(24)&0.9482 &0.1 &0.9999(239)&0.9634 &3.7 \\
1.3& 0.8706(10)& 0.8703(15)& 0.8737(33)&0.8698(19)&0.8706 &0.1 &0.8806(112)&0.8763 &0.5 \\
1.4& 0.8235(08)& 0.8217(14)& 0.8228(21)&0.8215(17)&0.8233 &0.2 &0.8250(71) &0.8210 &0.5 \\
1.5& 0.7973(08)& 0.7951(15)& 0.7933(28)&0.7953(18)&0.7977 &0.3 &0.7899(92) &0.7884 &0.2 \\
1.6& 0.7882(09)& 0.7861(17)& 0.7880(29)&0.7858(20)&0.7900 &0.5 &0.7918(95) &0.7735 &2.3 \\
1.7& 0.7947(12)& 0.7932(20)& 0.7860(45)&0.7943(24)&0.7992 &0.6 &0.7715(167)&0.7746 &0.4 \\
1.8& 0.8171(16)& 0.8172(24)& 0.8072(41)&0.8188(29)&0.8264 &0.9 &0.7872(178)&0.7917 &0.6 \\
1.9& 0.8585(24)& 0.8626(31)& 0.8538(98)&0.8640(40)&0.8757 &1.4 &0.8364(318)&0.8270 &1.1 \\
2.0& 0.9267(41)& 0.9408(43)& 0.9136(36)&0.9451(51)&0.9559 &1.1 &0.8591(355)& - & - \\
2.1& 1.0403(78)& 1.0816(67)& 1.0086(90)&1.0932(83)&1.0846 &0.8 &0.8625(927)& - & - \\
\hline
\multicolumn{10}{|l|}{$\beta=2.9$, $a^{-1}$=2.9 GeV, $L^*/a=11.6$}\\
\hline
1.1& 0.9945(12)& 1.0002(20)& 0.9916(23)&0.9922(21)&0.9909 &0.1 &0.9743(130)&0.9983 &2.5 \\
1.2& 0.9124(07)& 0.9147(12)& 0.9128(16)&0.9129(15)&0.9091 &0.4 &0.9089(58) &0.9110 &0.2 \\
1.3& 0.8625(07)& 0.8634(09)& 0.8583(11)&0.8587(10)&0.8579 &0.1 &0.8482(72) &0.8558 &0.9 \\
1.4& 0.8332(08)& 0.8336(12)& 0.8290(17)&0.8294(16)&0.8289 &0.1 &0.8199(79) &0.8236 &0.5 \\
1.5& 0.8223(08)& 0.8223(15)& 0.8178(15)&0.8181(14)&0.8179 &0.02&0.8088(75) &0.8098 &0.1 \\
1.6& 0.8266(07)& 0.8279(10)& 0.8231(14)&0.8234(13)&0.8234 &0.0 &0.8134(75) &0.8125 &0.1 \\
1.7& 0.8483(08)& 0.8519(13)& 0.8455(10)&0.8459(10)&0.8461 &0.03&0.8327(87) &0.8321 &0.1 \\
1.8& 0.8878(14)& 0.8986(17)& 0.8928(16)&0.8932(15)&0.8892 &0.5 &0.8811(91) &0.8711 &1.1 \\
1.9& 0.9572(23)& 0.9764(20)& 0.9569(21)&0.9582(19)&0.9591 &0.1 &0.9177(246)&0.9353 &1.9 \\
2.0& 1.0691(48)& 1.1059(24)& 1.0654(36)&1.0682(34)&1.0691 &0.1 &0.9846(499)&1.0358 &5.2 \\
\hline
\pagebreak
\multicolumn{10}{|l|}{$\beta=3.2$, $a^{-1}$=4.3 GeV, $L^*/a=17.2$}\\
\hline
1.1& 0.9649(97)& 0.9663(12)& 0.9618(13)&0.9591(24) &0.9591 &0.0  &0.9529(71)&0.9600 &0.7 \\
1.2& 0.9026(05)& 0.9025(11)& 0.8991(17)&0.8971(28) &0.8969 &0.02 &0.8924(68)&0.8951 &0.3 \\
1.3& 0.8635(06)& 0.8631(09)& 0.8603(11)&0.8587(19) &0.8595 &0.1  &0.8549(49)&0.8559 &0.1 \\
1.4& 0.8453(07)& 0.8460(11)& 0.8423(12)&0.8400(22) &0.8415 &0.2  &0.8348(62)&0.8365 &0.2 \\
1.5& 0.8455(05)& 0.8469(09)& 0.8421(11)&0.8392(21) &0.8404 &0.1  &0.8326(68)&0.8346 &0.2 \\
1.6& 0.8603(07)& 0.8626(10)& 0.8596(10)&0.8578(18) &0.8562 &0.2  &0.8535(51)&0.8497 &0.4 \\
1.7& 0.8939(08)& 0.9013(14)& 0.8943(11)&0.8901(23) &0.8908 &0.1  &0.8804(93)&0.8838 &0.4 \\
1.8& 0.9476(16)& 0.9652(13)& 0.9583(16)&0.9542(29) &0.9492 &0.5  &0.9447(97)&0.9417 &0.3 \\
1.9& 1.0351(25)& 1.0708(18)& 1.0518(22)&1.0403(52) &1.0408 &0.04&1.0138(240)&1.0329 &1.9 \\
2.0& 1.1747(62)& 1.2589(30)& 1.2033(33)&1.1697(121)&1.1850 &1.3 &1.0920(677)&1.1763 &7.7 \\
\hline
\multicolumn{10}{|l|}{$\beta=3.6$, $a^{-1}$=6.8 GeV, $L^*/a=27.2$}\\
\hline
1.0& 1.0210(07)& 1.0222(11)& 1.0204(16)&1.0183(36)&1.0162 &0.2 &1.0166(57)&1.0165 &0.01\\
1.1& 0.9467(05)& 0.9463(08)& 0.9418(09)&0.9369(29)&0.9407 &0.4 &0.9329(62)&0.9389 &0.6 \\
1.2& 0.8983(05)& 0.8975(09)& 0.8960(09)&0.8944(22)&0.8937 &0.1 &0.8931(36)&0.8907 &0.3 \\
1.3& 0.8730(05)& 0.8729(07)& 0.8696(09)&0.8660(26)&0.8681 &0.2 &0.8631(49)&0.8644 &0.2 \\
1.4& 0.8655(05)& 0.8660(07)& 0.8637(09)&0.8610(23)&0.8604 &0.1 &0.8589(41)&0.8566 &0.3 \\
1.5& 0.8750(05)& 0.8781(08)& 0.8738(10)&0.8690(31)&0.8697 &0.1 &0.8652(62)&0.8662 &0.1 \\
1.6& 0.9007(05)& 0.9075(07)& 0.9044(10)&0.9010(27)&0.8970 &0.4 &0.8983(50)&0.8943 &0.4 \\
1.7& 0.9459(08)& 0.9609(09)& 0.9547(11)&0.9479(39)&0.9462 &0.2 &0.9424(83)&0.9449 &0.3 \\
\hline
\pagebreak
\multicolumn{10}{|l|}{$\beta=4.1$, $a^{-1}$=12 GeV, $L^*/a=48$}\\
\hline
1.0& 0.9981(05)& 0.9972(08)& 0.9954(08)&0.9928(28)&0.9920 &0.1 &0.9919(37)&0.9904 &0.2 \\
1.1& 0.9381(04)& 0.9366(06)& 0.9329(07)&0.9273(34)&0.9324 &0.6 &0.9254(51)&0.9297 &0.5 \\
1.2& 0.9025(04)& 0.9016(06)& 0.8979(06)&0.8925(33)&0.8972 &0.5 &0.8907(49)&0.8941 &0.4 \\
1.3& 0.8867(04)& 0.8865(06)& 0.8831(07)&0.8781(31)&0.8815 &0.4 &0.8764(46)&0.8785 &0.2 \\
1.4& 0.8880(03)& 0.8896(05)& 0.8868(07)&0.8824(30)&0.8834 &0.1 &0.8810(43)&0.8808 &0.02\\
1.5& 0.9067(05)& 0.9107(09)& 0.9076(09)&0.9029(35)&0.9030 &0.02&0.9014(50)&0.9014 &0.0 \\
1.6& 0.9428(06)& 0.9531(09)& 0.9493(08)&0.9436(38)&0.9429 &0.1 &0.9417(55)&0.9429 &0.1 \\
1.7& 1.0003(12)& 1.0237(11)& 1.0193(12)&1.0127(48)&1.0087 &0.4 &1.0104(68)&1.0114 &0.1 \\
\hline
\multicolumn{10}{|l|}{$\beta=4.7$, $a^{-1}$=23 GeV, $L^*/a=92$}\\
\hline
0.9& 1.0618(06)& 1.0626(10)& 1.0614(11)&1.0592(36)&1.0554 &0.4 &1.0589(41)&1.0543 &0.4 \\
1.0& 0.9854(06)& 0.9836(09)& 0.9810(09)&0.9766(37)&0.9784 &0.2 &0.9760(44)&0.9762 &0.02\\
1.1& 0.9364(05)& 0.9351(08)& 0.9306(05)&0.9229(47)&0.9305 &0.8 &0.9217(58)&0.9278 &0.7 \\
1.2& 0.9100(06)& 0.9096(07)& 0.9054(05)&0.8981(45)&0.9048 &0.7 &0.8970(55)&0.9020 &0.6 \\
1.3& 0.9011(05)& 0.9015(08)& 0.9002(06)&0.8980(26)&0.8977 &0.04&0.8976(30)&0.8953 &0.3 \\
1.4& 0.9119(05)& 0.9143(07)& 0.9111(10)&0.9057(41)&0.9083 &0.3 &0.9048(49)&0.9067 &0.2 \\
1.5& 0.9394(05)& 0.9466(07)& 0.9419(08)&0.9337(52)&0.9381 &0.5 &0.9325(63)&0.9377 &0.6 \\
1.6& 0.9862(11)& 1.0010(12)& 0.9981(06)&0.9929(40)&0.9909 &0.2 &0.9922(48)&0.9926 &0.04\\
1.7& 1.0542(19)& 1.0903(15)& 1.0844(14)&1.0742(73)&1.0748 &0.1 &1.0727(87)&1.0799 &0.7 \\
\hline
\pagebreak
\multicolumn{10}{|l|}{$\beta=6.4$, $a^{-1}$=154 GeV, $L^*/a=616$}\\
\hline
0.8& 1.1139(05)& 1.1146(08)& 1.1116(07)&1.1057(43)&1.1081 &0.2 &1.1056(44)&1.1072 &0.1 \\
0.9& 1.0288(05)& 1.0248(07)& 1.0238(06)&1.0217(25)&1.0219 &0.02&1.0217(26)&1.0203 &0.1 \\
1.0& 0.9740(04)& 0.9711(05)& 0.9696(04)&0.9666(24)&0.9678 &0.1 &0.9666(24)&0.9658 &0.1 \\
1.1& 0.9426(03)& 0.9405(04)& 0.9381(03)&0.9334(30)&0.9375 &0.4 &0.9333(31)&0.9356 &0.2 \\
1.2& 0.9311(03)& 0.9300(06)& 0.9277(94)&0.9233(31)&0.9271 &0.4 &0.9232(32)&0.9254 &0.2 \\
1.3& 0.9385(03)& 0.9396(05)& 0.9382(04)&0.9354(23)&0.9352 &0.02&0.9353(23)&0.9339 &0.1 \\
1.4& 0.9624(03)& 0.9674(05)& 0.9665(05)&0.9647(20)&0.9629 &0.2 &0.9647(21)&0.9623 &0.2 \\
1.5& 1.0056(06)& 1.0194(07)& 1.0169(07)&1.0119(39)&1.0136 &0.2 &1.0118(40)&1.0143 &0.3 \\
1.6& 1.0746(12)& 1.1034(09)& 1.1005(11)&1.0950(48)&1.0950 &0.0 &1.0949(50)&1.0978 &0.3 \\
\hline    
\multicolumn{10}{|l|}{$\beta=8.85$, $a^{-1}$=2523 GeV, $L^*/a=10092$}\\
\hline
0.6& 1.3125(17)& 1.3578(14)& 1.3554(05)&1.3507(43)&1.3485 &0.2 &1.3507(43)&1.3495 &0.1 \\
0.7& 1.1812(06)& 1.1870(09)& 1.1862(07)&1.1846(29)&1.1822 &0.2 &1.1846(29)&1.1820 &0.2 \\
0.8& 1.0839(05)& 1.0798(09)& 1.0787(09)&1.0764(34)&1.0764 &0.0 &1.0764(34)&1.0755 &0.1 \\
0.9& 1.0175(03)& 1.0122(05)& 1.0101(03)&1.0060(28)&1.0089 &0.3 &1.0060(28)&1.0078 &0.2 \\
1.0& 0.9745(03)& 0.9710(05)& 0.9701(03)&0.9684(18)&0.9688 &0.04&0.9684(18)&0.9675 &0.1 \\
1.1& 0.9545(05)& 0.9522(07)& 0.9514(03)&0.9499(19)&0.9503 &0.04&0.9499(19)&0.9490 &0.1 \\
1.2& 0.9553(02)& 0.9549(04)& 0.9525(03)&0.9476(31)&0.9510 &0.4 &0.9476(31)&0.9500 &0.3 \\
1.3& 0.9719(05)& 0.9739(08)& 0.9728(05)&0.9706(24)&0.9712 &0.1 &0.9706(24)&0.9705 &0.01\\
1.4& 1.0066(03)& 1.0155(04)& 1.0160(05)&1.0169(18)&1.0133 &0.4 &1.0169(18)&1.0132 &0.4 \\
1.5& 1.0631(09)& 1.0888(07)& 1.0864(04)&1.0816(34)&1.0833 &0.2 &1.0816(34)&1.0841 &0.3 \\
1.6& 1.1480(25)& 1.1988(10)& 1.1974(06)&1.1946(32)&1.1928 &0.1 &1.1946(32)&1.1953 &0.1 \\
1.7& 1.2848(44)& 1.3802(12)&1.3705(09)&1.3513(118)&1.3651 &1.0&1.3513(118)&1.3705 &1.4 \\
\hline
\pagebreak                                               
\multicolumn{10}{|l|}{$\beta=21.0$, $a^{-1}=3.6\times 10^9$ GeV, 
$L^*/a=1.44\times 10^{10}$}\\
\hline
0.4& 1.5109(34)& 1.7583(07)&1.7420(06)&1.7095(197)&1.73370 &1.4 &1.7095(197)&1.7347 &1.5 \\
0.5& 1.3605(17)& 1.4332(06)& 1.4284(03)&1.4176(65)&1.42541 &0.6 &1.4176(65) &1.4257 &0.6 \\
0.6& 1.2319(05)& 1.2439(04)& 1.2430(02)&1.2412(16)&1.24099 &0.02&1.2412(16) &1.2410 &0.02\\
0.7& 1.1323(01)& 1.1263(03)& 1.1254(03)&1.1234(16)&1.12412 &0.1 &1.1234(16) &1.1239 &0.05\\
0.8& 1.0599(03)& 1.0518(02)& 1.0510(02)&1.0494(12)&1.04943 &0.0 &1.0494(12) &1.0492 &0.02\\
0.9& 1.0131(03)& 1.0062(02)& 1.0049(01)&1.0025(16)&1.00425 &0.2 &1.0025(16) &1.0040 &0.1 \\
1.0& 0.9892(02)& 0.9837(02)& 0.9829(02)&0.9814(11)&0.98212 &0.1 &0.9814(11) &0.9818 &0.04\\
1.1& 0.9851(02)& 0.9817(02)& 0.9813(02)&0.9805(9) &0.98019 &0.03&0.9805(09) &0.9799 &0.1 \\
1.2& 1.0004(02)& 0.9999(02)& 0.9994(01)&0.9985(8) &0.99821 &0.03&0.9985(08) &0.9980 &0.1 \\
1.3& 1.0339(02)& 1.0399(02)& 1.0393(03)&1.0382(11)&1.03848 &0.03&1.0382(11) &1.0384 &0.02\\
1.4& 1.0865(04)& 1.1072(03)& 1.1076(01)&1.1084(9) &1.10660 &0.2 &1.1084(09) &1.1066 &0.2 \\
1.5& 1.1600(09)& 1.2144(03)& 1.2148(04)&1.2154(14)&1.21365 &0.1 &1.2154(14) &1.2140 &0.1 \\
1.6& 1.2708(31)& 1.3827(04)& 1.3822(03)&1.3814(13)&1.38158 &0.01&1.3814(13) &1.3824 &0.1 \\
1.7& 1.4563(47)& 1.6713(09)&1.6600(05)&1.6375(137)&1.65782 &1.2 &1.6375(137)&1.6597 &1.4 \\
\hline
\end{longtable}

\begin{table}
  \leavevmode
\begin{center}
\caption{Fit parameters of $Z_V$ as a function of $M$
for the Plaquette action. }
\label{tab:fitM-P}                                  
\begin{tabular}{|c|ll|lll|c|}
\hline
$\beta$ & $M_c$ & $A_2$ & $B_0$ & $B_1$ & $B_2$& $\delta^{\rm max}$ (\%)\\
\hline
\multicolumn{7}{|c|}{ $8^3\times 16\times 16$} \\
\hline
5.8 & 1.758(14)& -1.441(64)& 0.7253(19)& -0.121(30)& -0.263(65)& 1.4 \\ 
6.0 & 1.661(47)& -1.449(71)& 0.7592(55)& -0.088(89)& -0.313(44)& 0.2 \\
6.2 & 1.658(12)& -0.957(46)& 0.7762(09)&  0.013(22)&  0.095(48)& 0.2 \\
6.5 & 1.562(13)& -1.290(14)& 0.7985(08)& -0.051(25)& -0.231(11)& 0.2 \\ 
6.8 & 1.530(14)&-1.478(120)& 0.8141(06)& -0.023(25)&-0.416(111)& 0.3 \\
7.4 & 1.483(08)& -1.124(76)& 0.8394(06)&  0.021(13)& -0.108(76)& 0.2 \\
8.0 & 1.399(01)& -1.066(04)& 0.8573(05)& -0.039(01)& -0.033(04)& 0.2 \\
9.6 & 1.319(07)& -1.026(53)& 0.8889(05)& -0.015(14)& 0.0002(589)& 0.1 \\
12.0& 1.242(01)& -1.083(19)& 0.9167(03)& -0.010(01)& -0.077(24)& 0.2 \\
24.0& 1.110(01)& -1.058(05)& 0.9628(02)& -0.008(02)& -0.068(08)& 0.1 \\ 
\hline
\multicolumn{7}{|c|}{ $4^3\times 8\times 16$} \\
\hline
5.8 & 1.639(11)& -1.570(23)& 0.7676(40)& -0.131(26)& -0.267(51)& 0.9 \\
\hline
\multicolumn{7}{|c|}{ $12^3\times 24\times 16$} \\
\hline
6.0 & 1.734(14)&-0.899(190)& 0.7441(17)& 0.0016(251)& 0.162(183)& 1.0 \\
6.2 & 1.902(13)&-0.552(03) & 0.8228(07)& 0.4180(15) & 0.389(03) & 0.5 \\
6.5 & 1.607(07)&-1.149(28) & 0.7946(09)& 0.0217(127)&-0.119(31) & 0.3 \\
6.8 & 1.645(71)&-0.591(128)& 0.8191(17)& 0.1823(121)& 0.391(121)& 0.3 \\ 
7.4 & 1.486(38)&-0.847(224)& 0.8366(11)& 0.0221(721)& 0.158(211)& 0.2 \\
8.0 & 1.455(08)&-0.931(02) & 0.8556(05)& 0.0670(14) & 0.076(05) & 0.1 \\
9.6 & 1.333(06)&-0.905(57) & 0.8871(04)& 0.0132(134)& 0.108(62) & 0.1 \\
12.0& 1.252(01)&-1.035(01) & 0.9153(02)& 0.0103(05) &-0.036(01) & 0.1 \\
24.0& 1.115(01)&-1.008(04) & 0.9617(01)& 0.0042(18) &-0.008(06) & 0.03\\
\hline
\end{tabular}                                         
\end{center}                                          
\end{table}                                           

\begin{table}
  \leavevmode
\begin{center}
\caption{Fit parameters of $Z_V$ as a function of $M$
for the RG action.}
\label{tab:fitM-RG}                                  
\begin{tabular}{|c|ll|lll|c|}
\hline
$\beta$ & $M_c$ & $A_2$ & $B_0$ & $B_1$ & $B_2$& $\delta^{\rm max}$ (\%)\\
\hline
\multicolumn{7}{|c|}{ $8^3\times 16\times 16$} \\
\hline
2.2 & 1.814(13)& -1.773(75) & 0.7125(21)& -0.111(28)& -0.614(98) & 1.6 \\
2.4 & 1.661(15)& -1.469(136)& 0.7596(40)& -0.092(28)& -0.394(140)& 0.7 \\
2.6 & 1.577(05)& -1.238(26) & 0.7864(09)& -0.053(11)& -0.168(35) & 0.3 \\
2.9 & 1.491(09)& -1.192(03) & 0.8226(10)& -0.042(16)& -0.142(10) & 0.1 \\
3.2 & 1.453(06)& -1.318(03) & 0.8443(03)&  0.005(01)& -0.282(03) & 0.1 \\
3.6 & 1.406(01)& -0.921(81) & 0.8664(04)&  0.035(02)&  0.090(82) & 0.1 \\
4.1 & 1.338(01)& -1.104(04) & 0.8856(02)&  0.008(01)& -0.100(04) & 0.01 \\
4.7 & 1.290(03)& -0.980(64) & 0.9023(04)&  0.008(07)&  0.031(69) & 0.1 \\
6.4 & 1.213(41)& -1.080(41) & 0.9306(03)&  0.018(03)& -0.083(45) & 0.1 \\
8.85& 1.143(01)& -1.058(01) & 0.9513(01)& -0.005(01)& -0.068(01) & 0.1 \\ 
21.0& 1.062(01)& -1.046(02) & 0.9806(01)&  0.002(01)& -0.060(04) & 0.1 \\
\hline
\multicolumn{7}{|c|}{ $4^3\times 8\times 16$} \\
\hline
2.2 & 1.675(08)& -1.856(141)& 0.7618(40)& -0.154(16)& -0.641(144)& 3.0 \\
2.4 & 1.583(16)& -1.556(58) & 0.7908(38)& -0.116(40)& -0.377(70) & 1.8 \\
\hline
\multicolumn{7}{|c|}{ $12^3\times 24\times 16$} \\
\hline
2.6 & 1.678(07)& -1.017(14)& 0.7863(21)& 0.098(12)& -0.006(26)& 0.6 \\ 
2.9 & 1.522(01)& -0.834(84)& 0.8170(11)& 0.008(03)&  0.177(91)& 0.3 \\
3.2 & 1.456(02)& -1.001(12)& 0.8401(05)& 0.005(04)&  0.012(13)& 0.2 \\
3.6 & 1.419(08)& -0.993(03)& 0.8643(05)& 0.055(01)&  0.004(06)& 0.1 \\
4.1 & 1.360(06)& -1.253(65)& 0.8832(07)& 0.050(11)& -0.245(68)& 0.03 \\ 
4.7 & 1.314(02)& -1.092(52)& 0.9000(04)& 0.055(04)& -0.085(56)& 0.1 \\
6.4 & 1.226(02)& -0.893(20)& 0.9288(03)& 0.044(03)&  0.120(23)& 0.1 \\
8.85& 1.151(01)& -1.003(09)& 0.9496(02)& 0.012(03)&  0.0002(118)& 0.03\\
21.0& 1.062(01)& -1.009(01)& 0.9797(01)& 0.004(01)& -0.011(02)& 0.1 \\
\hline
\end{tabular}                                         
\end{center}                                          
\end{table}                                           

\begin{table}
  \leavevmode
\begin{center}
\caption{Fit parameters of $Z_V$ as a function of $M$ and $g^2$.}
\label{tab:fit-g}                                  
\begin{tabular}{|cc|ll|ll|}
\hline
& & \multicolumn{2}{c|}{Plaquette}&\multicolumn{2}{c|}{RG}\\
\hline
& & $L=L^*$ & $L=\infty$ & $L=L^*$ & $L=\infty$ \\
\hline
 $M_c$ & $a_1$ & -0.5241   & -0.9026  & -0.6533$\times 10^{-1}$   & -0.3094 \\
       & $a_2$ & -0.1388   & -0.2365  & -0.3147$\times 10^{-2}$  & -0.5493 \\
       & $a_3$ &  0.5579$\times 10^{-1}$  & -0.1050  &  0.3841$\times 10^{-2}$  & -0.2635$\times 10^{-4}$ \\
\hline  
 $B_0$ & $a_4$ & -0.1166   & -0.9145  &  0.1378    & -0.3247 \\
       & $a_5$ &  0.8692$\times 10^{-2}$ &  0.7845$\times 10^{-1}$ & -0.1046    &  0.1959$\times 10^{-1}$ \\
       & $a_6$ & -0.7698$\times 10^{-1}$  &  0.4397$\times 10^{-1}$ & -0.4000$\times 10^{-2}$  &  0.8526$\times 10^{-3}$ \\
\hline
 $A_2$ & $a_7$ & -0.6437   & 0        & -0.3460    &  0 \\
       & $a_8$ & -0.2754   & 0        &  0         &  0 \\
       & $a_{9}$ &-0.3689 &-0.01092  & -0.2235$\times 10^{-1}$   &  0 \\
       & $a_{10}$ & 0.1002 & 0        &  0.1019$\times 10^{-1}$   &  0 \\
\hline
 $B_2$ & $a_{11}$ &-0.9481 & 0        & -0.3613    &  0 \\
\hline
\end{tabular}                                         
\end{center}                                          
\end{table}                                           

\clearpage

\end{document}